\documentclass[aps,prb,showpacs]{revtex4}%
\usepackage{amsmath}
\usepackage{graphicx}
\usepackage{amsfonts}%
\usepackage{amssymb}%

\begin{document}

\title{Surface scattering analysis of phonon transport in the quantum limit 
using  an elastic model}
\author{D.\ H.\ Santamore}
\author{M.\ C.\ Cross} \affiliation{Department of Physics,
California Institute of Technology 114-36, Pasadena, CA 91125}
\date{\today }

\begin{abstract}
We have investigated the effect on phonon energy transport in mesoscopic
systems and the reduction in the thermal conductance in the quantum limit due
to phonon scattering by surface roughness using full 3-dimensional elasticity
theory for an elastic beam with a rectangular cross-section. At low
frequencies we find power laws for the scattering coefficients that are
strongly mode dependent, and different from the $\omega^{2}$ dependence,
deriving from Rayleigh scattering of scalar waves, that is often assumed. The
scattering gives contributions to the reduction in thermal conductance with
the same power laws. At higher frequencies the scattering coefficients becomes
large at the onset frequency of each mode due to the flat dispersion here. We
use our results to attempt a quantitative understanding of the suppression of
the thermal conductance from the universal value observed in experiment.

\end{abstract}
\maketitle

\section{Introduction}

Landauer's formulation of quantum transport showed that when elastic
scattering dominates, the electrical conductance can be related to the
transmission coefficient of the electron waves\cite{L57}. In the ideal case of
no scattering, this leads to a universal conductance that is quantized in
units of $e^{2}/h$ at low temperatures, with an additional quantum of
conductance added as each channel or mode of the conductance pathway opens up.
The application of similar ideas to the phonon counterpart, namely thermal
conductance, was recently derived by a number of authors\cite{ACR98, RK98,
B99}, and is now recognized\cite{BV00} to be related to earlier work on the
entropy transport at low temperatures\cite{P83}. Rego and Kirczenow have
extended the concept of the universality of the thermal conductance to
particles of arbitrary statistics (anyons)\cite{RK99}.

In the case of electrical resistance, the chemical potential or
the number of conducting modes can be varied at very low
temperatures, giving sharp jumps between various quantized values
of the resistance. On the other hand, thermal transport by phonons
necessarily requires nonzero temperatures to populate the modes of
the conducting pathway, and the width of the Bose distribution
function smears out the quantization of the conductance. Only at
very low temperatures, where just the modes of the conducting
pathway with zero frequency at long wavelengths contribute to the
thermal conductance, the quantization of the ideal conductance
becomes apparent in a universal thermal conductance $N_{0}K_{u}$
with $K_{u}=(\pi^{2}/3)k_{B}^{2}T/h$ the universal conductance per
mode, with $k_{B}$ Boltzmann's constant and $h$ Planck's constant,
and $N_{0}$ is the number of modes with zero frequency at long
wavelengths, which is four for a freely suspended elastic beam
connecting the two thermal reservoirs. Note that this value of the
low temperature conductance in the absence of scattering is
independent of the dimensions and elastic properties of the
thermal pathway.

A low temperature thermal conductance consistent with the
predicted universal value was measured by Schwab et
al.\cite{SHWR00} in experiments on a lithographically defined
mesoscopic suspended beam (of dimensions about
$1\mathrm{\mu m}\times200\mathrm{nm}\times60\mathrm{nm}$). Whilst
their elegant experiment displays the universality of ballistic
phonon transport, the experiment also showed a \emph{decrease} in
the thermal conductance below the universal value in the
temperature range of $0.08\mathrm{K}<T<0.4\mathrm{K}$ that cannot
be explained by the ballistic theory, since in this theory an
increase in the thermal conductance is expected as the temperature
is raised and more modes are excited. The decrease in thermal
conductance is presumably associated with the scattering of the
thermal phonons, and can be understood using the ideas of Landauer
in terms of the scattering coefficient of the vibrational waves.
This is the topic of the present paper.

In this paper we calculate the effect on the low temperature thermal
conductance of the scattering of the thermal phonons by surface roughness,
which is likely to be the major source of scattering in mesoscopic samples.
The scattering of scalar waves, described by the simple wave equation, in
waveguides with rough surfaces has been investigated by many workers,
including ourselves, using both numerical and analytic
methods\cite{S97,KFFL99,SFMY99,B99,SC00}. However, for the low frequency modes
of interest in the low temperature thermal conductance, the physical
vibrational waves have quite different properties than the waves in the scalar
model. For example the dispersion relations of the modes are different, with
two of the four modes with zero long-wavelength frequency having a quadratic
dispersion at small wave vectors, rather than the linear dependence given by
the simple scalar theory. To understand the experimental results
quantitatively, a more accurate treatment of the vibrational waves is needed.
At low temperatures the wavelengths of the thermally excited modes are large
compared with the atomic spacing, and so a treatment based on the equations of
macroscopic elasticity theory is appropriate. Blencowe\cite{B95} has
considered the scattering of elastic waves in a thin plate waveguide with
rough surfaces, but prior to our work, the scattering of elastic waves
confined in a beam-like wave guide with rough surfaces has not been considered.

Previously, we have investigated the effect of surface scattering
on the low temperature thermal conductance using the scalar wave
model\cite{SC00}. In that paper we noted the apparent discrepancy
between the results of scalar model with a simple assumption for
the nature of the surface roughness and the data by Schwab et al.
below a temperature of $0.1\mathrm{K}$: the data seemed to show a
delay of the onset of scattering as the temperature increased that
was not predicted by the model. However, since the scalar model
does not properly account for the properties of the elastic waves,
it was not clear whether this discrepancy is due to an inadequate
modelling of the surface roughness, or the flaw in the description
of the waves themselves. To resolve this matter, and obtain a more
accurate account of the scattering of the waves by rough surfaces,
we develop a theory of based on the full elasticity equations, and
use this to calculate the thermal conductance at low temperatures.
A short version of this work has been previously
published\cite{SC01l}.

In Section\ \ref{Sec_Formalism}, the scattering of elastic waves confined to a
beam of rectangular cross-section with rough surfaces is calculated using the
full three dimensional elasticity theory. We use a Green theorem approach, and
calculate the scattering coefficient to quadratic order in the amplitude of
the surface roughness. These results are quite general, but rather intractable
for further progress, since the structure of the modes in an elastic beam
cannot be determined in closed form. Thus in Section\ \ref{Sec_plate} we
reduce the expressions to a thin plate limit to provide a closed form for the
displacement fields, and to obtain analytical expressions for the scattering
behavior. In Section \ref{Sec_scattering} the general behavior of the
scattering and the effect on the thermal conductance is analyzed in detail,
using a simple description of the surface roughness, to investigate the
physical consequences of the novel features of the elastic waves. In Section
\ref{Sec_experiment} we use our theory to attempt to fit the data of Schwab et
al\cite{SHWR00} using more realistic descriptions of the surface roughness. A
number of the more difficult issues that arise in the elasticity theory are
described in appendices.

Although our main interest is the scattering of thermally excited vibrational
waves in mesoscopic systems at low temperatures, the formulation of the
surface scattering is quite general, and can be applied to other situations,
such as the scattering of mechanically excited modes in macroscopic samples
for example.

\section{General Formalism}

\label{Sec_Formalism}

\subsection{The model}

The main focus of this paper is the effect of surface roughness on the low
temperature thermal conductance of mesoscopic structures. The geometry we
consider is a freely suspended elastic beam, which we call the bridge,
connecting two thermal reservoirs. We will consider a beam of rectangular
cross-section of dimensions width $W$ (in the $y$ direction) and depth $d$ (in
the $z$ direction). Mesoscopic structures are often produced lithographically
from epitaxially grown material. We choose a convention that the depth is the
dimension in the growth direction, and the width in the lithographically
defined transverse direction. We define the length of the rectangular beam of
nominally uniform cross section as $L$. In practice the bridge may be joined
to the reservoirs smoothly, by a portion of continuously growing width, to
eliminate or reduce the scattering of the vibration modes off a sharp
junction. We will suppose that the scattering by roughness is important only
in some narrower portion of length $L$.

The thermal conductance is given by the expression\cite{ACR98, RK98, B99}%

\begin{equation}
K={\frac{\hbar^{2}}{{k_{B}T^{2}}}}\sum_{m}{\frac{1}{{2\pi}}}\int_{\omega_{m}%
}^{\infty}\mathcal{T}_{m}(\omega){\frac{\omega^{2}e^{\beta\hbar\omega}%
}{{(e^{\beta\hbar\omega}-1)^{2}}}}d\omega, \label{eq:conductance}%
\end{equation}
where $\omega_{m}$ is the cutoff frequency of the $m$ th mode, $\beta
=1/(k_{B}T)$, and $T$ is the temperature. The integration is over the
frequency $\omega$ of the modes $m$ that propagate in the structure. The
transmission coefficient is unity for the ideal case. Any scattering reduces
the thermal conductance, and scattering of the lowest modes can reduce the
conductance below the universal value at low temperatures.

\begin{figure}[ptb]
\begin{center}
\includegraphics[ height=2.8037in, width=4.0612in ]{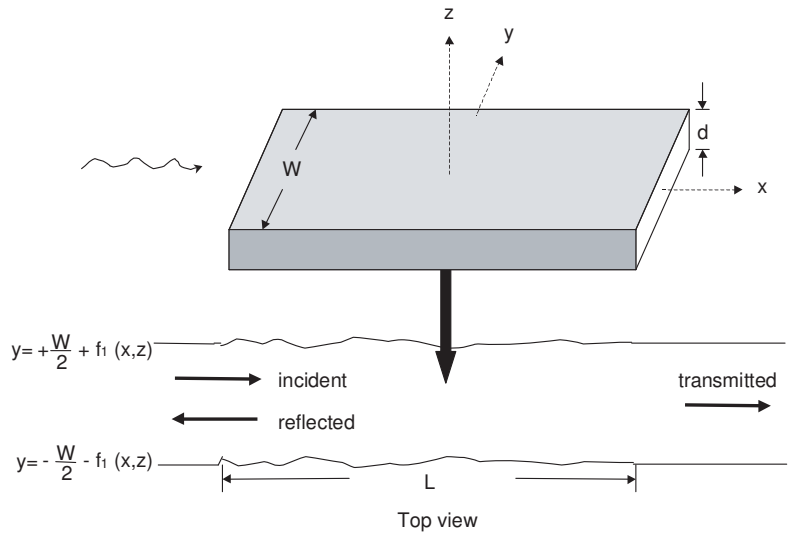}
\end{center}
\caption{Top: Three dimensional elastic beam with rectangular cross-section.
The rough surfaces are on the top, bottom, and sides. Bottom: Side view of the
mathematical model of the structure actually used for the scattering
calculation.}%
\label{modelstructure}%
\end{figure}

To actually perform the scattering calculation we imbed the rough beam of
length $L$ in an infinite beam of the same cross section but with smooth
surfaces outside of the region of length $L$, Fig.\ \ref{modelstructure}. Thus
the mathematical calculation is the scattering of a wave incident from
$x=-\infty$ on a rough portion of the beam with surfaces at $y=\pm W/2\pm
f_{1}\left(  x,z\right)  $ and at $z=\pm d/2\pm f_{2}\left(  x,y\right)  $,
with the height functions $f_{1,2}$, defining the roughness, nonzero only in a
finite region $0<x<L$. Forward scattering is evaluated from the intensity of
waves as $x\rightarrow+\infty$, and backward scattering from the intensity of
waves as $x\rightarrow-\infty$.

To calculate the scattering amplitude, we take a Green function approach
similar to our previous work on the scalar model\cite{SC00}.

The displacement field $\mathbf{u}$ away from any sources satisfies the wave
equation:
\begin{equation}
\rho\partial_{t}^{2}u_{i}=\partial_{j}T_{ij} \label{wave}%
\end{equation}
where $\rho$ is the mass density, and%
\begin{equation}
T_{ij}=C_{ijkl}\partial_{k}u_{l} \label{stress strain}%
\end{equation}
is the stress tensor field with $C_{ijkl}$ the elastic modulus tensor. The
subscript $i$ runs over the $3$ Cartesian coordinates, we use the symbol
$\partial_{x}$ to denote the derivative $\partial/\partial x$ etc., and
repeated indices are to be summed over. The displacement field satisfies
stress free boundary conditions at the surfaces%
\begin{equation}
T_{ij}n_{j}|_{S}=0 \label{free-boundary}%
\end{equation}
where $S$ denotes the surface boundaries and $n_{j}$ is normal to the surface.
Assuming harmonic time dependence at frequency $\omega$, Eq.\ (\ref{wave})
becomes
\begin{equation}
\rho\omega^{2}u_{i}+C_{ijkl}\partial_{j}\partial_{k}u_{l}=0.
\label{helmholtz-u}%
\end{equation}

We approximate the material of the system as an isotropic solid. Then the
elastic modulus tensor is
\begin{equation}
C_{ijkl}=\lambda\delta_{ij}\delta_{kl}+\mu\left(  \delta_{ik}\delta
_{ji}+\delta_{il}\delta_{kj}\right)  \label{elasticity-tensor}%
\end{equation}
where $\lambda$\ and $\mu$\ are Lam\'{e} constants ($\mu$ is also the shear modulus)%
\begin{equation}
\lambda = E \sigma / (1 + \sigma)(1 - 2\sigma)
,\quad\mu=E/2\left(  1+\sigma\right)
\label{Lame}%
\end{equation}
with $E$ Young's modulus and $\sigma$\ the Poisson ratio.

Even in a rectangular beam geometry the displacement fields in the propagating
modes yielded by these equations are complicated, and cannot be found
analytically. The modes can be grouped into four classes according to their
signature under the parity operations $y\rightarrow-y$ and $z\rightarrow-z$.
Some modes show regions of anomalous dispersion where the group velocity
$d\omega/dk$ is negative: these regions require a careful examination of the
notions of ``forward'' and ``backward'' scattering for the waves. The lowest
frequency mode of each class has a frequency that tends to zero at small wave
number. These four modes are the only ones excited at low enough temperature,
and are the ones contributing to the universal thermal conductance. The
structure of these modes at small wave numbers is simple and can be calculated
using familiar macroscopic arguments of elasticity theory: they are the
compression, torsion, and (two orthogonal) bending modes.

We define a Green function $G_{iq}(\mathbf{x},\mathbf{x}^{\prime};t,t^{\prime
})$ to satisfy the wave equation with a source term $-\delta_{iq}%
\delta(\mathbf{x}-\mathbf{x}^{\prime})\delta(t-t^{\prime})$, and $\Gamma
_{ijq}$ to be the corresponding stress%
\begin{equation}
\Gamma_{ijq}\equiv C_{ijkl}\partial_{k}G_{lq}. \label{gamma}%
\end{equation}
It is convenient to introduce the frequency space version of the Green
function
\begin{equation}
G_{iq}(\mathbf{x};\mathbf{x}^{\prime};t,t^{\prime})=\int\,\frac{d\omega}{2\pi
}G_{iq}(\mathbf{x};\mathbf{x}^{\prime};\omega)e^{-i\omega\left(  t-t^{\prime
}\right)  }, \label{full-Green}%
\end{equation}
with a similar expression defining $\Gamma_{ijq}(\mathbf{x},\mathbf{x}%
^{\prime};\omega)$. Inserting $G$, $\Gamma$, and the source term into
Eq.\ (\ref{helmholtz-u}) gives%
\begin{equation}
\rho\omega^{2}G_{iq}(\mathbf{x},\mathbf{x}^{\prime};\omega)+\partial_{j}%
\Gamma_{ijq}(\mathbf{x},\mathbf{x}^{\prime};\omega)=-\delta_{iq}\delta\left(
\mathbf{x}-\mathbf{x}^{\prime}\right)  \label{helmholtz-G}%
\end{equation}
where $\mathbf{x}$ is the observation coordinate and $\mathbf{x}^{\prime}$ is
the source coordinate.

Equations\ (\ref{helmholtz-u}) and (\ref{helmholtz-G}) lead to Green's theorem
expressing the displacement field at frequency $\omega$ in terms of a surface
integral%
\begin{equation}
u_{q}(\mathbf{x})=\int_{S^{\prime}}\left[  n_{j}^{\prime}T_{ij}\left(
\mathbf{x}^{\prime}\right)  G_{iq}\left(  \mathbf{x}^{\prime},\mathbf{x;\omega
}\right)  -n_{j}^{\prime}u_{i}\left(  \mathbf{x}^{\prime}\right)  \Gamma
_{ijq}\left(  \mathbf{x}^{\prime},\mathbf{x;\omega}\right)  \right]
dS^{\prime}. \label{total field}%
\end{equation}
We are free to choose any closed integration surface $S^{\prime}$. One choice
is to use the physical rough surface thereby eliminating the first term in
Eq.\ (\ref{total field}) due to the boundary condition
Eq.\ (\ref{free-boundary}). However, the resulting integration over the rough
surface is not easy. Instead, we integrate over the smoothed surfaces at
$y=\pm W/2$ and $z=\pm d/2$ and impose the boundary conditions on the Green
function to be stress free on these smoothed surfaces%
\begin{equation}
\left.  \Gamma_{ijq}n_{j}\right|  _{S}=0, \label{Eq_Green-bc}%
\end{equation}
together with cross sections at $x^{\prime}\rightarrow\pm\infty$ to close the surface.

The total field $\mathbf{u}$ can be written as the sum of incident and
scattered waves
\begin{equation}
\mathbf{u}=\mathbf{u}^{\mathrm{in}}+\mathbf{u}^{\mathrm{sc}}\text{.}%
\end{equation}
It can be shown (see Appendix \ref{Appendix_Separate}) that the integration
over the sections at $x^{\prime}\rightarrow\pm\infty$ on the right hand side
of Eq.\ (\ref{total field}) just gives $u_{q}^{\mathrm{in}}$. In the
integration over the smoothed surfaces at $y=\pm W/2$ and $z=\pm d/2$ the
second term in the integrand vanishes due to Eq.\ (\ref{Eq_Green-bc}). Thus we
find the expression for the scattered field%

\begin{equation}
u_{q}^{\mathrm{sc}}(\mathbf{x})=\int_{S}\left[  n_{j}^{\prime}T_{ij}\left(
\mathbf{x}^{\prime}\right)  G_{iq}\left(  \mathbf{x}^{\prime},\mathbf{x;\omega
}\right)  \right]  dS^{\prime}, \label{scatterd-field-w/bd}%
\end{equation}
with the surface $S$ the smoothed surfaces $y=\pm W/2$ and $z=\pm d/2$. The
stress field $T_{ij}$ on the smoothed surface is evaluated by expanding about
its value on the \emph{rough} surfaces, where Eq.\ (\ref{free-boundary}) applies.

The rest of the section goes as follows: firstly, we find an explicit
expression for the Green function with stress free boundary conditions; then
we apply the boundary perturbation method to project the stress at the rough
surfaces onto the smooth surfaces by expanding the stress free boundary terms
around the smooth surfaces using the small roughness as the expansion
parameter; and finally we evaluate the strength of the scattered waves to give
the scattering coefficient.

\subsection{Green function}

We evaluate $G_{iq}(\mathbf{x},\mathbf{x}^{\prime};\omega)$ as an expansion in
the complete orthonormal set of normal modes $\mathbf{u}^{\left(  k,m\right)
}\left(  \mathbf{x}\right)  $ in the ideal geometry, which satisfy
Eq.\ (\ref{helmholtz-u}) and stress free boundaries at the smooth surfaces.
Here $k$ is the wave number in the $x$ direction, and $m$ labels the branch of
the dispersion curve. We define $\omega_{m}(k)$ as the frequency of the mode
$m$ at wave number $k$ in the ideal geometry. The modes satisfy the
completeness relation%
\begin{equation}
\sum_{m}\int\,\frac{dk}{2\pi}u_{i}^{(k,m)}(\mathbf{x}^{\prime})^{\ast}%
u_{j}^{(k,m)}(\mathbf{x})=\delta_{ij}\delta(\mathbf{x}-\mathbf{x}^{\prime}).
\label{completeness}%
\end{equation}
Substituting this expression on the right hand side of Eq.\ (\ref{helmholtz-G}%
) leads to the expression for the Green function%
\begin{equation}
G_{iq}(\mathbf{x}^{\prime},\mathbf{x};\omega)=-\sum_{m}\frac{1}{2\pi}%
\int_{-\infty}^{\infty}dk\,\,\frac{\phi_{i}^{(k,m)}(y^{\prime},z^{\prime
})^{\ast}\phi_{q}^{(k,m)}(y,z)}{\rho\left[  \left(  \omega+i\epsilon\right)
^{2}-\omega_{m}^{2}(k)\right]  }e^{ik\left(  x-x^{\prime}\right)  },
\label{GreenFT}%
\end{equation}
where we write%
\begin{equation}
u_{i}^{(k,m)}(\mathbf{x})=\phi_{i}^{(k,m)}(y,z)e^{ikx} \label{Eq_u-phi}%
\end{equation}
with$\ \phi_{i}^{(k,m)}$\ giving the transverse dependence\ of the
displacement field. In Eq.\ (\ref{GreenFT}) $\epsilon$ is a positive
infinitesimal number to incorporate causality, $G_{iq}(\mathbf{x}%
,\mathbf{x}^{\prime};t,t^{\prime})=0$ for $t<t^{\prime}$.

Equation\ (\ref{GreenFT}) can now be evaluated by contour integration. The
integrand has poles labelled by an index $n$ near values $k=k_{n}$ on the real
axis which are given by solutions to the dispersion relation $\omega_{m}%
(k_{n})=\omega$ for all branches $m$. (We take an incident wave with
$\omega>0$.) Note that for branches with regions of anomalous dispersion there
may be more than one solution to this equation for some $\omega$, so that the
index $n$ is not identical to the branch index $m$. The poles are shifted
slightly off the real axis by the infinitesimal $\varepsilon$ in
Eq.\ (\ref{GreenFT}), and are given by expanding about $k_{n}$%
\[
k=k_{n}+\frac{i\epsilon}{\omega_{n}v_{g}^{\left(  n\right)  }},
\]
with $v_{g}^{\left(  n\right)  }$\ the group velocity at the $n$th pole
$\left.  d\omega_{m}/dk\right|  _{k=k_{n}}$. Notice the poles are in the upper
half plane for $v_{g}^{(n)}>0$, and in the lower half plane for $v_{g}%
^{\left(  n\right)  }<0$.

Now we can perform the $k$ integration by complex integration. Consider first
the case, $x>x^{\prime}$. The contour must be closed in the upper half plane
so that the contribution from the semicircle at large $|k|$ vanishes. The
contour integration then picks up contributions from the poles in the upper
half plane, i.e. wave numbers with $v_{g}^{\left(  n\right)  }>0$. On the
other hand, for $x<x^{\prime}$ the contour must be closed in the lower half
plane and it is poles at wave numbers with $v_{g}^{\left(  n\right)  }<0$ that
give nonzero residue. Forward scattering or backscattering is thus seen to be
determined by the sign of the group velocity $v_{g}^{\left(  n\right)  }$
rather than by the sign of $k_{n}$, as indeed would be expected physically.

Evaluating the residues gives the expression for the Green function:
\begin{equation}
G_{iq}(\mathbf{x}^{\prime},\mathbf{x};\omega)=i\left.  \sum_{n}\right.
^{\prime}\,\,\frac{u_{i}^{(n)}(\mathbf{x}^{\prime})^{\ast}u_{q}^{(n)}%
(\mathbf{x})}{2\rho\omega_{n}\,v_{g}^{\left(  n\right)  }},
\label{Green function}%
\end{equation}
where $u_{i}^{(n)}(\mathbf{x})$ is written for $u_{i}^{(k,m)}(\mathbf{x})$\ at
the value of the wave number $k=k_{n}$ satisfying $\omega_{m}(k_{n})=\omega$.
The prime on the sum is used to denote the fact that the sum runs over
intersections $n$ with $v_{g}^{(n)}>0$ for $x>x^{\prime}$, and over $n$ with
$v_{g}^{(n)}<0$ for $x<x^{\prime}$.

The group velocity $v_{g}^{\left(  n\right)  }$ does not have an analytical
expression for a rectangular beam, and is obtained by differentiating the
dispersion curve which must be found numerically. Alternatively, to avoid
numerical differentiation, we can rewrite $v_{g}^{\left(  n\right)  }$ in
terms of the average power flow in mode $n$. Since $u_{i}^{\left(  n\right)
}$\ is normalized, the power $P_{n}$ in mode $n$ can be written as
\begin{equation}
P_{n}=\frac{1}{2}\operatorname{Re}\int\int\left(  -i\omega T_{ix}^{\left(
n\right)  }u_{i}^{\left(  n\right)  \ast}\right)  dydz=\frac{1}{2}\rho
\omega^{2}v_{g}^{\left(  n\right)  }, \label{power}%
\end{equation}
the first expression of the equality expressing the energy flux in terms of
the rate of work done across a section, and the second in terms of the group
velocity and the average energy density evaluated as twice the average kinetic
energy. Then $v_{g}^{\left(  n\right)  }$ can be evaluated in terms of $P_{n}$
as
\begin{equation}
v_{g}^{\left(  n\right)  }=2P_{n}/\rho\omega^{2} \label{vg-Power}%
\end{equation}
and $P_{n}$ has an expression directly in terms of displacement field given by
the first equality in Eq.\ (\ref{power})%
\begin{equation}
P_{n}=\frac{1}{2}\operatorname{Re}\int\int\left(  -i\omega T_{ix}^{\left(
n\right)  }u_{i}^{\left(  n\right)  \ast}\right)  dydz. \label{Power_integral}%
\end{equation}
This expression for $v_{g}^{\left(  n\right)  }$ can also be derived directly
from the equations of motion \cite{A}.

\subsection{Boundary perturbation}

In this section we show the boundary perturbation technique for the rough
surfaces on the sides (i.e.\ the $xz$ boundary planes). We work out the
scattering coefficient explicitly for the surface near $y=W/2$. The surface
near $y=-W/2$ will give a similar contribution and, assuming uncorrelated
roughness on the two surfaces, is accounted for by multiplying the
single-surface scattering rate by $2$ at the end of the calculation. The
results for the top and bottom surfaces can be obtained by interchanging $y$
and $z$ whenever they occur in the indices in the displacement fields and
stress tensors in the calculation below.

In order to calculate the stress on the smooth surface appearing in
Eq.\ (\ref{scatterd-field-w/bd}), we expand the stress $T_{ij}$\ in a Taylor
series about the flat surface, and impose stress free boundary conditions at
the rough surface which is the small distance $f_{1}$ away. We also assume
$f_{1}$\ is differentiable.

The unit vector $\hat{n}$ normal to the rough boundaries to first order in
$f_{1}$ is
\begin{equation}
\hat{n}\simeq\hat{y}-\partial_{x}f_{1}\left(  x,z\right)  \hat{x}-\partial
_{z}f_{1}\left(  x,z\right)  \hat{z}.
\end{equation}
Then the stress free surface boundary conditions Eq.\ (\ref{free-boundary})
can be written%
\begin{equation}
\left[  T_{iy}-\partial_{x}f_{1}\left(  x,z\right)  T_{ix}-\partial_{z}%
f_{1}\left(  x,z\right)  T_{iz}\right]  _{y=\frac{W}{2}+f}=0
\label{bondary-sides}%
\end{equation}
Now expanding Eq.\ (\ref{bondary-sides}) in the neighborhood of $y=W/2$ and
taking only the lowest order in $f_{1}$ and $f_{1}^{\prime}$, we obtain%
\begin{align}
\left.  T_{xy}\right|  _{y=\frac{W}{2}}  &  \simeq\left.  \left(  \partial
_{x}f_{1}\left(  x,z\right)  T_{xx}+\partial_{z}f_{1}\left(  x,z\right)
T_{xz}-f_{1}\left(  x,z\right)  \partial_{y}T_{xy}\right)  \right|
_{y=\frac{W}{2}}\label{xz-perturbed1}\\
\left.  T_{zy}\right|  _{y=\frac{W}{2}}  &  \simeq\left.  \left(  \partial
_{x}f_{1}\left(  x,z\right)  T_{zx}+\partial_{z}f_{1}\left(  x,z\right)
T_{zz}-f_{1}\left(  x,z\right)  \partial_{y}T_{zy}\right)  \right|
_{y=\frac{W}{2}}\label{xz-perturbed2}\\
\left.  T_{yy}\right|  _{y=\frac{W}{2}}  &  \simeq\left.  -f_{1}\left(
x,z\right)  \partial_{y}T_{yy}\right|  _{y=\frac{W}{2}} \label{xz-perturbed3}%
\end{align}
where the first two expressions for $T_{xy}$ and $T_{zy}$ have been used to
simplify $T_{yy}$. Since the terms on the right hand side of
Eqs.\ (\ref{xz-perturbed1}-\ref{xz-perturbed3}) are explicitly first order in
the small parameter $f_{1}$, the stress field $T_{ij}$ on the right hand side
can be evaluated at zeroth order, i.e. for ideal smooth surfaces. These
results are used in Eq.\ (\ref{scatterd-field-w/bd}).

\subsection{Scattering coefficient}

We now evaluate the expression for the scattered field given by an integration
over the beam surfaces Eq. (\ref{scatterd-field-w/bd}). To calculate the
scattering coefficient, we consider an incident wave of unit amplitude in a
single mode $m$. Again in this section we will outline the calculation for the
scattering by the single surface at $y=W/2$, and will include the effects of
the other surfaces at the end. We therefore have%
\begin{equation}
u_{q}^{\text{sc}}\left(  \mathbf{x}\right)  =\int\int\left[  T_{iy}\left(
\mathbf{x}^{\prime}\right)  G_{iq}\left(  \mathbf{x}^{\prime},\mathbf{x}%
;\omega\right)  \right]  _{y^{\prime}=\frac{W}{2}}dx^{\prime}dz^{\prime}.
\label{sc field}%
\end{equation}

We can now evaluate the forward and backscattering amplitudes by using
Eq.\ (\ref{Green function}) for the Green function in Eq.\ (\ref{sc field}),
and evaluating the scattered wave at large positive and negative $x$%
\begin{equation}
u_{q}^{\mathrm{sc}}\left(  x\rightarrow\infty,y,z\right)  \simeq\int_{-\infty
}^{x}dx^{\prime}\int_{-\frac{d}{2}}^{\frac{d}{2}}dz^{\prime}\sum
_{n,v_{g}^{(n)}>0}\frac{i}{2\rho\omega\,v_{g}^{\left(  n\right)  }}\left[
T_{iy}\left(  \mathbf{x}^{\prime}\right)  u_{i}^{\left(  n\right)  }\left(
\mathbf{x}^{\prime}\right)  ^{\ast}\right]  _{y^{\prime}=\frac{W}{2}}%
u_{q}^{\left(  n\right)  }\left(  \mathbf{x}\right)  , \label{x-inf}%
\end{equation}%
\begin{equation}
u_{q}^{\mathrm{sc}}\left(  x\rightarrow-\infty,y,z\right)  \simeq\int
_{x}^{\infty}dx^{\prime}\int_{-\frac{d}{2}}^{\frac{d}{2}}dz^{\prime}%
\sum_{n,v_{g}^{(n)}<0}\frac{i}{2\rho\omega\,v_{g}^{\left(  n\right)  }}\left[
T_{iy}\left(  \mathbf{x}^{\prime}\right)  u_{i}^{\left(  n\right)  }\left(
\mathbf{x}^{\prime}\right)  ^{\ast}\right]  _{y^{\prime}=\frac{W}{2}}%
u_{q}^{\left(  n\right)  }\left(  \mathbf{x}\right)  . \label{x-minf}%
\end{equation}
The stress tensor $T_{ij}$ corresponding to the full displacement field of the
wave is evaluated from Eqs.\ (\ref{xz-perturbed1}-\ref{xz-perturbed3}). Since
these expressions explicitly include the small roughness amplitude $f_{1}$ on
the right hand side, to calculate the scattering at lowest order in the
roughness amplitude it is sufficient to replace all $T_{ij}$ on the right hand
side by the value $T_{ij}^{(m)}$ in the incident mode $m$. From
Eqs.\ (\ref{x-inf}), (\ref{x-minf}) we see that $\mathbf{u}^{\text{\textrm{sc}%
}}(\mathbf{x)}$ is expressed as a sum over modes $\mathbf{u}^{\left(
n\right)  }(\mathbf{x)}$, and the coefficient of each mode is then the
scattering amplitude $t_{n,m}$ from incident mode $m$ into mode $n$, so that
\begin{equation}
t_{n,m}=\int_{-\infty}^{\infty}dx\int_{-\frac{d}{2}}^{\frac{d}{2}}%
dz\frac{i}{2\rho\omega\,v_{g}^{\left(  n\right)  }}\left[  T_{iy}^{(m)}\left(
\mathbf{x}\right)  u_{i}^{\left(  n\right)  }\left(  \mathbf{x}\right)
^{\ast}\right]  _{y=W/2}, \label{scat-amplitude}%
\end{equation}
where we can now extend the integration limit to $\pm\infty$ since $f_{1}$,
and so the integrand, is zero outside the domain of roughness $0<x<L$. Again
mode indices $n$ for which $v_{g}^{(n)}>0$ represent the forward-scattered
waves and those with $v_{g}^{(n)}<0$ the backward-scattered waves.

Now use the expression for the stress tensor on the smooth surfaces obtained
in the previous section Eqs.\ (\ref{xz-perturbed1}-\ref{xz-perturbed3}) and
integrate the resulting expressions by parts with respect to $x$ or $z$ to
rewrite the terms in $\partial_{x}f_{1}$ and $\partial_{z}f_{1}$ as
integrations over $f_{1}$. After these manipulations we find $t_{n,m}$ can be
written
\begin{equation}
t_{n,m}=-\frac{i}{2\rho\omega v_{g}^{\left(  n\right)  }}\int_{-\infty
}^{\infty}dx\int_{-\frac{d}{2}}^{\frac{d}{2}}dz\,f_{1}\left(  x,z\right)
\Gamma^{\left(  m,n\right)  }\left(  x,z\right)
\end{equation}
where%
\begin{align}
\Gamma^{\left(  m,n\right)  }(x,z)  &  =\left[  \left(  \partial_{x}%
T_{xx}^{\left(  m\right)  }+\partial_{y}T_{xy}^{\left(  m\right)  }%
+\partial_{z}T_{xz}^{\left(  m\right)  }\right)  u_{x}^{\left(  n\right)
\ast}+\left(  \partial_{x}T_{zx}^{\left(  m\right)  }+\partial_{y}%
T_{zy}^{\left(  m\right)  }+\partial_{z}T_{zz}^{\left(  m\right)  }\right)
u_{z}^{\left(  n\right)  \ast}\right. \nonumber\\
&  \left.  +\partial_{y}T_{yy}^{\left(  m\right)  }u_{y}^{\left(  n\right)
\ast}+T_{xx}^{\left(  m\right)  }\partial_{x}u_{x}^{\left(  n\right)  \ast
}+T_{zz}^{\left(  m\right)  }\partial_{z}u_{z}^{\left(  n\right)  \ast}%
+T_{zx}^{\left(  m\right)  }\left(  \partial_{x}u_{z}^{\left(  n\right)  \ast
}+\partial_{z}u_{x}^{\left(  n\right)  \ast}\right)  \right]  _{y=W/2}.
\label{gamma-1}%
\end{align}
Applying the equations of motion Eq.\ (\ref{helmholtz-u}) and
remembering$\left.  T_{iy}^{(m)}\right|  _{y=W/2}=0$ for all $i$ and for all
$x,z$ leads to the somewhat simpler expression%
\begin{align}
\Gamma^{\left(  m,n\right)  }(x,z\mathbf{)}  &  =\left[  -\rho\omega
^{2}\left(  u_{x}^{(m)}u_{x}^{\left(  n\right)  \ast}+u_{y}^{(m)}%
u_{y}^{\left(  n\right)  \ast}+u_{z}^{(m)}u_{z}^{\left(  n\right)  \ast
}\right)  \right. \nonumber\\
&  \left.  +T_{xx}^{\left(  m\right)  }\partial_{x}u_{x}^{\left(  n\right)
\ast}+T_{zz}^{\left(  m\right)  }\partial_{z}u_{z}^{\left(  n\right)  \ast
}+T_{xz}^{\left(  m\right)  }\left(  \partial_{z}u_{x}^{\left(  n\right)
\ast}+\partial_{x}u_{z}^{\left(  n\right)  \ast}\right)  \right]  _{y=W/2}.
\end{align}
Notice that the scattering separates into a kinetic term (the first line)\ and
a stress term (the second line).

The above form for $\Gamma^{(m,n)}$ is still neither instructive nor practical
for numerical evaluation. It can be further simplified using the expressions
Eqs.\ (\ref{stress strain}) and (\ref{Lame}) for the stress tensor in terms of
the displacements. First we use the boundary condition $T_{yy}^{(m)}=0$ for
the y-stress to give at $y=W/2$%
\begin{equation}
\partial_{y}u_{y}^{(m)}=-\frac{\sigma}{\left(  1-\sigma\right)  }(\partial
_{x}u_{x}^{(m)}+\partial_{z}u_{z}^{\left(  m\right)  }). \label{bc1}%
\end{equation}
This can be used to simplify the expressions for the other components of the
stress tensors at $y=W/2$%
\begin{align}
T_{xx}^{(m)}  &  =\frac{E}{\left(  1-\sigma^{2}\right)  }(\partial_{x}%
u_{x}^{(m)}+\sigma\partial_{z}u_{z}^{\left(  m\right)  }),\label{bc2}\\
T_{zz}^{(m)}  &  =\frac{E}{\left(  1-\sigma^{2}\right)  }(\sigma\partial
_{x}u_{x}^{(m)}+\partial_{z}u_{z}^{\left(  m\right)  }),\label{bc3}\\
T_{xz}^{(m)}  &  =\frac{E}{2\left(  1+\sigma\right)  }(\partial_{x}u_{z}%
^{(m)}+\partial_{z}u_{x}^{(m)}). \label{bc4}%
\end{align}
Inverting these gives at $y=W/2$%
\begin{align}
\partial_{x}u_{x}^{\left(  m\right)  }  &  =\frac{1}{E}(T_{xx}^{\left(
m\right)  }-\sigma T_{zz}^{\left(  m\right)  }),\label{bc5}\\
\partial_{z}u_{z}^{\left(  m\right)  }  &  =\frac{1}{E}(T_{zz}^{\left(
m\right)  }-\sigma T_{xx}^{\left(  m\right)  }),\label{bc6}\\
\partial_{x}u_{z}^{(m)}+\partial_{z}u_{x}^{(m)}  &  =\frac{2(1+\sigma)}%
{E}T_{xz}^{\left(  m\right)  }. \label{bc7}%
\end{align}
We emphasize that Eqs.\ (\ref{bc1}-\ref{bc7}) are only true on the stress free
boundaries, and are not generally true in the bulk of the material.

Using these results we get%
\begin{equation}
t_{n,m}=-\frac{i}{2\rho\omega\,v_{g}^{\left(  n\right)  }}\int_{-\frac{d}{2}%
}^{\frac{d}{2}}dz\,\tilde{f}_{1}\left(  k_{m}-k_{n},z\right)  \bar{\Gamma
}^{\left(  m,n\right)  }(z) \label{scattered-field-final}%
\end{equation}
with%
\begin{align}
\bar{\Gamma}^{\left(  m,n\right)  }  &  =\left\{  \rho\omega^{2}(\phi
_{x}^{(m)}\phi_{x}^{\left(  n\right)  \ast}+\phi_{y}^{(m)}\phi_{y}^{\left(
n\right)  \ast}+\phi_{z}^{(m)}\phi_{z}^{\left(  n\right)  \ast})\right.
\label{numerator}\\
&  \left.  -\frac{1}{E}\left[  (\bar{T}_{xx}^{\left(  m\right)  }\bar{T}%
_{zz}^{\left(  n\right)  \ast}+\bar{T}_{zz}^{\left(  m\right)  }\bar{T}%
_{zz}^{\left(  n\right)  \ast})-(\sigma\bar{T}_{zz}^{\left(  m\right)  }%
\bar{T}_{xx}^{\left(  n\right)  \ast}+\bar{T}_{xx}^{\left(  m\right)  }\bar
{T}_{zz}^{\left(  n\right)  \ast})\right]  -\frac{1}{\mu}\bar{T}_{xz}^{\left(
m\right)  }\bar{T}_{zx}^{\left(  n\right)  \ast}\right\}  _{y=W/2}\nonumber
\end{align}
where we have introduced the explicit $x$ dependence of $u_{i}^{(n)}%
(\mathbf{x})$ as in Eq.\ (\ref{Eq_u-phi}) and\ the stress tensor%
\begin{equation}
T_{ij}^{(n)}(\mathbf{x})=\bar{T}_{ij}(y,z)e^{ik_{n}x},
\end{equation}
so that the $x^{\prime}$ integration is just the Fourier transform $\tilde{f}$
of the roughness function, and $\bar{\Gamma}$ is a function of the $z$
coordinate only.

Alternatively, using Eqs.\ (\ref{bc2}-\ref{bc4}) we can derive an expression
explicitly in the displacement fields, which is useful for numerical
evaluation,%
\begin{align}
\bar{\Gamma}^{\left(  m,n\right)  }  &  =\left\{  \rho\omega^{2}(\phi
_{x}^{(m)}\phi_{x}^{\left(  n\right)  \ast}+\phi_{y}^{(m)}\phi_{y}^{\left(
n\right)  \ast}+\phi_{z}^{(m)}\phi_{z}^{\left(  n\right)  \ast})\right.
\nonumber\\
&  -\frac{2\mu}{\left(  1-\sigma\right)  }\left[  (k_{m}k_{n}\phi_{x}%
^{(m)}u_{x}^{\left(  n\right)  \ast}+\partial_{z}\phi_{z}^{\left(  m\right)
}\partial_{z}\phi_{z}^{\left(  n\right)  \ast})+\sigma(ik_{m}\phi_{x}%
^{(m)}\partial_{z}\phi_{z}^{\left(  n\right)  \ast}-ik_{n}\partial_{z}\phi
_{z}^{\left(  m\right)  }\phi_{x}^{\left(  n\right)  \ast})\right] \nonumber\\
&  \left.  -\mu(ik_{m}\phi_{z}^{(m)}\partial_{z}\phi_{x}^{\left(  n\right)
\ast}+k_{m}k_{n}\phi_{z}^{(m)}\phi_{z}^{\left(  n\right)  \ast}+\partial
_{z}\phi_{x}^{(m)}\partial_{z}\phi_{x}^{\left(  n\right)  \ast}-ik_{n}%
\partial_{z}\phi_{x}^{(m)}\phi_{z}^{\left(  n\right)  \ast})\right\}
_{y=W/2}. \label{gamma-U}%
\end{align}

The scattering rate is given by multiplying $\left|  t_{n,m}\right|  ^{2}$ by
the ratio of the group velocities in the scattered and incident
waves\footnote{Note that the scattering amplitude normalized by the energy
flux $\bar{t}_{n,m}=\sqrt{\left|  v_{g}^{\left(  n\right)  }\right|  /\left|
v_{g}^{\left(  m\right)  }\right|  }t_{n,m}$ can be seen from
Eq.\ (\ref{scattered-field-final}) to explicitly satisfy the reciprocity
relation $\bar{t}_{n,m}=\bar{t}_{-m,-n}^{\ast}$. For systems in which energy
is conserved, as in our case, reciprocity is equivalent to time-reversal
invariance\cite{P98}.}. We also treat the roughness of the surface
statistically, and take an ensemble average (denoted by angular brackets) to
give the final expression for the scattering rate $\gamma_{n,m}$ from mode $m$
to mode $n$ by the per unit length of single rough surface at $y=W/2$ given
by
\begin{equation}
\gamma_{n,m}L=\frac{v_{g}^{\left(  n\right)  }}{v_{g}^{\left(  m\right)  }%
}\left\langle \left|  t_{n,m}\right|  ^{2}\right\rangle =\frac{1}{4\rho
^{2}\omega^{2}v_{g}^{\left(  m\right)  }v_{g}^{\left(  n\right)  }%
}\left\langle \left|  \int_{-\frac{d}{2}}^{\frac{d}{2}}dz\tilde{f}_{1}\left(
k_{m}-k_{n},z\right)  \bar{\Gamma}^{\left(  m,n\right)  }(z)\right|
^{2}\right\rangle . \label{scattering-probabiity-nm}%
\end{equation}

We are interested in the reduction of the phonon heat transport due to rough
surfaces. Only the backscattered waves (those with $v_{g}^{\left(  n\right)
}<0$) reduce the amount of heat transmitted. Thus we define $\gamma_{m}$, the
thermal attenuation coefficient of mode $m$ per unit length, to be the sum of
the scattering rates from the incident mode $m$ to \emph{all} possible
backscattered modes, per unit length of rough surface. This can be written for
scattering off the single rough surface considered so far%
\begin{equation}
\gamma_{m}L=\sum\limits_{\substack{n\\v_{g}^{(n)}<0}}\gamma_{n,m}%
L=\sum_{\substack{n\\v_{g}^{(n)}<0}}\frac{1}{4\rho^{2}\omega^{2}v_{g}^{\left(
m\right)  }v_{g}^{\left(  n\right)  }}\left\langle \left|  \int_{-\frac{d}{2}%
}^{\frac{d}{2}}dz\tilde{f}_{1}\left(  k_{m}-k_{n},z\right)  \bar{\Gamma
}^{\left(  m,n\right)  }(z)\right|  ^{2}\right\rangle .
\label{scattering-probability}%
\end{equation}

To include the second rough side surface, assuming uncorrelated roughness, we
simply have to multiply the expression for $\gamma_{m}$ by a factor of $2$.
The expression for scattering off the top and bottom surfaces, if these are
rough too, can be derived in a similar manner, and the result may be obtained
by exchanging $y$ and $z$ in Eq.\ (\ref{scattering-probability}). The total
scattering rate is the sum of the scattering off all the surfaces.

We have assumed that the amplitude of the surface roughness is small, allowing
us to use perturbation theory to derive the above expressions. In this case,
the attenuation coefficient gives an exponential decay of the energy flux, so
that the energy flux transmission coefficient is%
\begin{equation}
\mathcal{T}_{m}=\exp\left[  -\gamma_{m}L\right]  .
\end{equation}

\section{Thin Plate Limit}

\label{Sec_plate}Although the expression in the previous section is general
and applicable to any rectangular waveguide with rough surfaces,
there are no closed-form expressions for the displacement fields
in general, and so a direct evaluation of the scattering would
have to be done completely numerically. Here, we instead use the
\emph{thin plate approximation} $d\ll W$ \cite{LF7,CL00}, which
yields closed form expressions for the displacement fields of the
modes (in terms of a dispersion curves $\omega_{m}(k)$ given by
numerical solution of a simple transcendental equation). The thin
plate limit captures the important properties of the elastic
modes, for example the quadratic dispersion of the bending modes
at small wave numbers, and regions of anomalous dispersion, as
well as providing analytical expressions enabling us to do further
analysis of the scattering. The thin plate theory is applicable
where the thickness of the sample is much less than the width and
the wavelengths are much greater than the thickness, which is the
case for many mesoscopic systems at low temperatures.

The use of the thin plate limit for mesoscopic structures was
proposed in reference \cite{CL00}, where the calculation of the
structure of the modes is described in more detail. It is found
that the modes can be separated into two classes: \emph{in-plane
modes}, where the polarization of the displacement is largely in
the $xy$ plane (together with small strains in the $z$-direction
given by the Poisson effect) and the displacement field is
completely specified by giving the \emph{vertically averaged
horizontal displacement components }$\bar{u}_{x}(x,y)$ and
$\bar{u}_{y}(x,y)$; and \emph{flexural modes}, where the
displacement is primarily in the $z$ direction and is specified by
a vertical displacement field $\bar{u}_{z}(x,y)$. Within each
class we can further distinguish the modes by their parity under
$y\rightarrow-y$. For the in-plane modes we define the mode as
even if $\bar{u}_{x}\left(  x,-y\right)  =\bar{u}_{x}\left(
x,y\right)  $\ and odd if $\bar{u}_{x}\left(  x,-y\right)
=-\bar{u}_{x}\left(  x,y\right)  $. Similarly, the even flexural
modes have $\bar{u}_{z}\left(  x,-y\right) =\bar{u}_{z}\left(
x,y\right)  $ and the odd modes have $\bar{u}_{z}\left(
x,-y\right)  =-\bar{u}_{z}\left(  x,y\right)  $. As in the general
case, there are four branches of the dispersion curves that tend
to zero frequency as the wave number goes to zero, corresponding
to one mode from each of these classes. The low frequency, even in
plane mode corresponds to the compression mode, and the odd mode
to a bending mode. The low frequency even, flexural mode
corresponds to the second bending mode, and the low frequency odd
flexural mode is the torsion mode.

Explicit expressions for the displacement fields can be obtained using the
method described in reference \cite{CL00}. For the in-plane modes we find, up
to a normalization factor $A_{1}$ that is common to both even and odd parity
waves, the even modes%

\begin{equation}
\bar{u}_{x}\left(  x,y\right)  =ikA_{1}\left[  \frac{k^{2}-\chi_{1}^{2}%
}{2k^{2}}\cos\left(  \frac{\chi_{2}W}{2}\right)  \cos\left(  \chi_{1}y\right)
-\cos\left(  \chi_{2}y\right)  \cos\left(  \frac{\chi_{1}W}{2}\right)
\right]  e^{ikx} \label{even-ux-filed}%
\end{equation}%
\begin{equation}
\bar{u}_{y}\left(  x,y\right)  =A_{1}\left[  \frac{k^{2}-\chi_{1}^{2}}%
{2\chi_{1}}\cos\left(  \frac{\chi_{2}W}{2}\right)  \sin\left(  \chi
_{1}y\right)  +\chi_{2}\cos\left(  \frac{\chi_{1}W}{2}\right)  \sin\left(
\chi_{2}y\right)  \right]  e^{ikx}, \label{even-uy-field}%
\end{equation}
and the odd modes%
\begin{equation}
\bar{u}_{x}\left(  x,y\right)  =ikA_{1}\left[  \frac{k^{2}-\chi_{1}^{2}%
}{2k^{2}}\sin\left(  \chi_{1}y\right)  \sin\left(  \frac{\chi_{2}W}{2}\right)
-\sin\left(  \frac{\chi_{1}W}{2}\right)  \sin\left(  \chi_{2}y\right)
\right]  e^{ikx} \label{odd-ux-field}%
\end{equation}%
\begin{equation}
\bar{u}_{y}\left(  x,y\right)  =-A_{1}\left[  \frac{k^{2}-\chi_{1}^{2}}%
{2\chi_{1}}\cos\left(  \chi_{1}y\right)  \sin\left(  \frac{\chi_{2}W}%
{2}\right)  +\chi_{2}\sin\left(  \frac{\chi_{1}W}{2}\right)  \cos\left(
\chi_{2}y\right)  \right]  e^{ikx}, \label{odd-uy-field}%
\end{equation}
where $\chi_{1}=\left(  \omega^{2}/c_{t}^{2}-k^{2}\right)  ^{1/2}$ and
$\chi_{2}=\left(  \omega^{2}/c_{l}^{2}-k^{2}\right)  ^{1/2}$,\ with $c_{t}%
$\ the transverse sound velocity and $c_{l}$\ the longitudinal
velocity \emph{in a large thin plate}
\begin{equation}
c_{t} = \sqrt{\frac{E}{2\rho \left( 1+\sigma \right) }},\quad
c_{l} = \sqrt{\frac{E}{\rho \left( 1-\sigma ^{2}\right) }},
\end{equation}
and $\omega$ and $k$ related by the dispersion curve which must be
found numerically. In the thin plate limit it is sufficient to
take for the in-plane modes
\begin{align}
u_{x}(x,y,z)  &  \simeq\bar{u}_{x}(x,y),\\
u_{y}(x,y,z)  &  \simeq\bar{u}_{y}(x,y),\\
u_{z}(x,y,z)  &  \simeq0.
\end{align}

Similarly, the vertical displacement field for the even flexural modes is%
\begin{equation}
\bar{u}_{z}\left(  x,y\right)  =A_{2}\left[  \cosh\left(  \frac{\chi_{-}W}%
{2}\right)  \cosh\left(  \chi_{+}y\right)  -\frac{k^{2}\sigma-\chi_{+}^{2}%
}{k^{2}\sigma-\chi_{-}^{2}}\cosh\left(  \frac{\chi_{+}W}{2}\right)
\cosh\left(  \chi_{-}y\right)  \right]  e^{ikx}, \label{even_flex}%
\end{equation}
and for the odd flexural modes%
\begin{equation}
\bar{u}_{z}\left(  x,y\right)  =A_{2}\left[  \sinh\left(  \frac{\chi_{-}W}%
{2}\right)  \sinh\left(  \chi_{+}y\right)  -\frac{k^{2}\sigma-\chi_{+}^{2}%
}{k^{2}\sigma-\chi_{-}^{2}}\sinh\left(  \frac{\chi_{+}W}{2}\right)
\sinh\left(  \chi_{-}y\right)  \right]  e^{ikx}, \label{odd_flex}%
\end{equation}
where $\chi_{+}=\left(  k^{2}+\sqrt{\rho d/D}\omega\right)  ^{2}$ and
$\chi_{-}=\left(  k^{2}-\sqrt{\rho d/D}\omega\right)  ^{2}$, with
$D=Ed^{3}/12\left(  1-\sigma^{2}\right)  $ the flexural rigidify, and again
$\omega$ and $k$ are related by the appropriate dispersion curve. In the
classical thin plate theory, the displacement fields are given in terms of
$\bar{u}_{z}$ by the expressions%
\begin{align}
u_{x}(x,y,z)  &  \simeq-z\partial_{x}\bar{u}_{z}(x,y),\\
u_{y}(x,y,z)  &  \simeq-z\partial_{y}\bar{u}_{z}(x,y),\\
u_{z}(x,y,z)  &  \simeq\bar{u}_{z}(x,y).
\end{align}
This approximation is adequate for evaluating the surface stress integrals in
Eq.\ (\ref{scattering-probability}) but turns out not to be sufficiently
accurate to evaluate the energy flux expression for the group velocity
Eq.\ (\ref{Power_integral}). We discuss this case in section
\ref{Subsec_Power_Flexural} below.

\subsection{Ideal thermal conductance}%

\begin{figure}
[ptb]
\begin{center}
\includegraphics[
height=3.1522in,
width=3.8977in
]%
{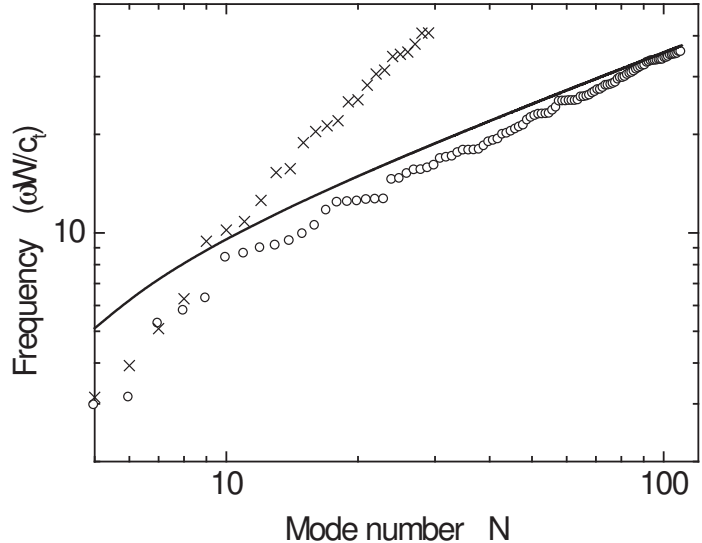}%
\caption{Mode frequency $\omega_{N}$ as a function of mode number $N$: crosses
- thin plate theory; circles - xyz algorithm; solid line - bulk mode density
of states calculation. A thickness to width ratio $d/W=0.38$ was used.}%
\label{cut-off}%
\end{center}
\end{figure}
Since our quantitative calculation of the scattering\ coefficient relies on
the analytic expressions for the elastic modes available only in the thin
plate limit, it is essential to estimate the temperature range where the thin
plate limit is applicable for a given experimental structure. On the other
hand, as the wavelength becomes much smaller than the dimensions of the
structure, we should to be able to treat the waves in terms of separate
longitudinal and transverse waves in the bulk of the material, without
worrying too much about the complicated standing wave transverse mode
structure important for the long wavelength modes. In this regime, which we
refer to as the bulk mode limit, the counting of the modes is insensitive to
the details of the boundary conditions, and is the same as for a scalar wave
approximation. The ideal thermal conductance depends only on cut-off frequency
of the modes (see Eq.\ (\ref{eq:conductance})), and we can assess the
applicability of these simple limiting approximations by comparing the mode
cutoff frequencies with results from a numerical calculation of the full
elasticity theory. For the full elastic theory, we use the ``xyz
algorithm''\cite{NAW97}.

For the bulk mode calculation, there are three polarizations (one
longitudinal and two transverse) with propagation velocities
$c_{3l}$ and $c_{t}$, respectively with $c_{t}$ as before and
\begin{equation}
c_{3l} = \sqrt{\frac{E\left( 1-\sigma \right) }{\rho \left(
1+\sigma \right) \left( 1-2\sigma \right) }}.
\end{equation}
The precise details of the boundary conditions are unimportant in
the mode counting for large mode numbers. If we assume standing
waves in the transverse direction corresponding to zero normal
derivative boundary
conditions on the wave functions, the cutoff frequencies are%
\begin{equation}
\omega_{t,mn}=c_{t}\sqrt{\left(  \frac{m\pi}{W}\right)  ^{2}+\left(
\frac{n\pi}{d}\right)  ^{2}},\text{\quad(two fold degenerate)}%
\end{equation}
for the transverse waves, and%
\begin{equation}
\omega_{l,mn}=c_{3l}\sqrt{\left(  \frac{m\pi}{W}\right)
^{2}+\left(
\frac{n\pi}{d}\right)  ^{2}},\text{\quad(non degenerate)}%
\end{equation}
for the longitudinal waves, with $m,n=0,1,2\ldots$. For large $m,n$ we can use
the continuous form for the frequency $\omega_{N}$ of the $N$th mode%
\begin{equation}
N=\frac{dW}{4\pi}\omega_{N}^{2}\left(  \frac{2}{c_{t}^{2}}+\frac{1}{c_{3l}^{2}%
}\right)  . \label{bulk_mode}%
\end{equation}

Figure \ref{cut-off} shows the cutoff frequencies as a function of mode number
for the thickness to width ratio $d/W=0.38$. The thin plate theory gives a
good approximation at lower frequencies. The accuracy of thin plate theory
becomes better as $d/W$ gets smaller. For example, in the case of $d/W=0.1$
(not shown), the error in the cutoff frequencies of the first $13$ modes is
less than $3\%$, whilst the error is as large as $5\%$ for the first $7$ modes
for the case $d/W=0.38$ shown in the figure. In terms of the ideal
(no-scattering) thermal conductance (Eq.\ (\ref{eq:conductance}) with the
transmission coefficient set to unity), we find that for $d/W=0.38$ the error
in the thermal conductance is less than $4\%$ up to $T\sim0.4$ K. Thus the
thin plate limit is adequate to examine the scattering effects in this
temperature range. At large frequencies $\omega W/c_{t}>30$ the elasticity
theory results approach closely the continuum bulk mode calculations,
Eq.\ (\ref{bulk_mode}). The thin plate approximation clearly fails in this
limit, since it predicts $N\propto\omega$ corresponding to a $2d$ structure.

\subsection{Attenuation coefficient in the thin plate limit}

The thin plate approximation is implemented by noticing that the
stress free boundary conditions imply that the stress components
$T_{iz}$ are zero on the top and bottom surfaces. For small
thickness this implies that the components $T_{iz}$ for any $i$
are small everywhere. In most situations these components can be
approximated as zero\cite{LF7}. This simplifies many of the terms
appearing in Eq.\ (\ref{scattered-field-final}). Also, at low
temperatures, only modes with no strong dependence on the $z$
coordinate will be excited, so that the mode sum extends over
modes with increasing numbers of nodes in the $y$ direction only.

In this section we calculate the scattering of the elastic waves
by surface roughness for a thin plate. We assume that the
roughness is confined to the sides, since in the experiments
theses are prepared lithographically, whereas the top and bottom
surfaces are produced by the epitaxial growth process.

For simplicity, we assume the roughness function $f_{1}$ has no
$z$ dependence. This is probably a reasonable description of the
roughness produced by a typical lithographic process of
anisotropic chemical etch\footnote{Without this assumption, we
would find slightly different $z$ averages of the roughness
function $\tilde{f}_{1}$ involved for the scattering of the
in-plane modes and of the flexural modes---in fact a direct
average for the inplane modes and an average weighted by $z^{2}$
for the flexural modes. In addition there would now be scattering
from in-plane to flexural modes, and \emph{vice versa}.}. Then the
Fourier transformed roughness function $\tilde{f}_{1}\left(
k_{m}-k_{n}\right)  $\ can be pulled outside of the $z$ integral
in Eq.\ (\ref{scattering-probability}) and the statistical average
over the roughness can be performed to give%
\begin{equation}
\left\langle \left|  \tilde{f}_{1}\left(  k\right)  \right|  ^{2}\right\rangle
=\tilde{g}\left(  k\right)  L,
\end{equation}
where $\tilde{g}\left(  k\right)  $ is the Fourier transform of the roughness
correlation function%
\[
\tilde{g}\left(  k\right)  =\int dxe^{-ikx}\left\langle f_{1}\left(  x\right)
f_{1}\left(  0\right)  \right\rangle .
\]

Equation (\ref{scattering-probability}) leads to the back-scattering rate from
mode $m$ to mode $n$
\begin{equation}
\gamma_{n,m}=\frac{\tilde{g}\left(  k_{m}-k_{n}\right)  }{2\rho^{2}\omega
^{2}v_{g}^{\left(  m\right)  }v_{g}^{\left(  n\right)  }}\left|
\int_{-\frac{d}{2}}^{\frac{d}{2}}dz\bar{\Gamma}^{\left(  m,n\right)
}(z)\right|  ^{2}. \label{attenuation-coefficient}%
\end{equation}
where Eq.\ (\ref{scattering-probability}) is multiplied by a factor of $2$ to
account for the two surfaces at $y=\pm W/2$.

With the closed forms of the displacement fields at hand, we can obtain the
analytical expression for the attenuation coefficient. We first evaluate
$\bar{\Gamma}^{(m,n)}$ from Eq.\ (\ref{numerator}). Since $T_{iz}^{(m)}%
\simeq0$, the expression for $\bar{\Gamma}$ reduces to%
\begin{equation}
\bar{\Gamma}^{\left(  m,n\right)  }\simeq\left[
\rho\omega^{2}\left( \phi_{x}^{(m)}\phi_{x}^{\left( n\right)
\ast}+\phi_{y}^{(m)}\phi _{y}^{\left(  n\right)
\ast}+\phi_{z}^{(m)}\phi_{z}^{\left(  n\right)  \ast
}\right)  -\frac{1}{E}\left(  \bar{T}_{xx}^{\left(  m\right)  }\bar{T}%
_{xx}^{\left(  n\right)  \ast}\right)  \right]  _{y=W/2}. \label{thin-A}%
\end{equation}
In addition, putting $T_{zz}^{(m)}$ in Eq.\ (\ref{bc4}) at the stress free
boundary to zero gives
\begin{equation}
\partial_{z}u_{z}^{\left(  m\right)  }=-\sigma\partial_{x}u_{x}^{(m)}%
\end{equation}
so that $T_{xx}^{(m)}$ from Eq.\ (\ref{bc2}) simplifies to%
\begin{equation}
T_{xx}^{(m)}=E\partial_{x}u_{x}^{(m)}.
\end{equation}
Now Eq.\ (\ref{attenuation-coefficient}) can be written as%
\begin{equation}
\gamma_{n,m}=\frac{\tilde{g}\left(  k_{m}-k_{n}\right)  }{2\rho^{2}\omega
^{2}v_{g}^{\left(  m\right)  }v_{g}^{\left(  n\right)  }}\left|
\int_{-\frac{d}{2}}^{\frac{d}{2}}dz\left[  \rho\omega^{2}\phi_{i}^{(m)}%
\phi_{i}^{\left(  n\right)  \ast}+Ek_{s}k_{n}\phi_{x}^{(m)}\phi_{x}^{(n)\ast
}\right]  _{y=\frac{W}{2}}\right|  ^{2}, \label{gamma-thin}%
\end{equation}
where the index $i$ is summed over $x,y,z$. The scattering in the thin plate
limit is seen to have two components: the kinetic term, the first term in the
$[]$ in Eq.\ (\ref{gamma-thin}), which involves all components of the
displacement; and the stress term, the second term, which just depends on the
longitudinal displacement.

To see how the scattering rate scales with the parameters it is useful to
rewrite Eq.\ (\ref{gamma-thin}) as%
\begin{multline}
\gamma_{n,m}L=\frac{\tilde{g}\left(  k_{m}-k_{n}\right)  L}{2W^{4}}%
\times\frac{W^{2}\omega^{2}}{v_{g}^{\left(  m\right)  }v_{g}^{\left(
n\right)  }}\label{gamma_thin_nonorm}\\
\times\frac{\left|  \int_{-d/2}^{d/2}dz\left[  \phi_{i}^{(m)}\phi_{i}^{\left(
n\right)  \ast}+\frac{Ek_{s}k_{n}}{\rho\omega^{2}}\phi_{x}^{(m)}\phi
_{x}^{(n)\ast}\right]  _{y=\frac{W}{2}}\right|  ^{2}}{\left(  \int
_{-d/2}^{d/2}dz\int_{-W/2}^{W/2}\frac{dy}{W}\phi_{i}^{(m)}\phi_{i}^{(m)\ast
}\right)  ^{1/2}\left(  \int_{-d/2}^{d/2}dz\int_{-W/2}^{W/2}\frac{dy}{W}%
\phi_{i}^{(n)}\phi_{i}^{(n)\ast}\right)  ^{1/2}}.
\end{multline}
The first factor is a dimensionless measure of the strength of the roughness;
the second factor is a dimensionless ratio that depends, through the
dispersion relation, only on the geometric ratio $d/W$ and the Poisson ratio
$\sigma$; and the final factor involves integrals over the displacement
fields, where we have introduced the explicit normalization factors in the
denominator so that we may evaluate the ratio using convenient unnormalized
expressions for the displacements.

\subsection{Evaluating the group velocity}

As we have seen in Eq.\ (\ref{vg-Power}), we can avoid evaluating the group
velocity appearing in Eq.\ (\ref{gamma-thin}) via numerical differentiating
the dispersion curve by instead relating the group velocity to the energy flux
in the mode, which in turn can be written as an explicit integral
Eq.\ (\ref{Power_integral}). Thus we need to evaluate the expression (we
suppress the mode index in this section )%
\begin{equation}
P=-\frac{1}{2}\operatorname{Re}\left[  i\omega\int\int\left(  T_{xx}%
u_{x}^{\ast}+T_{yx}u_{y}^{\ast}+T_{zx}u_{z}^{\ast}\right)  dydz\right]  .
\label{Power_integral_ex}%
\end{equation}
involving the displacement fields and their derivatives.

In the thin plate limit the $z$ components of the stress are small. If we
approximate $T_{zz}=0$ then expressions Eqs.\ (\ref{stress strain}) and
(\ref{Lame}) can be used to evaluate the $z$-component of the strain%
\begin{equation}
\partial_{z}u_{z}=-\frac{\sigma}{(1-\sigma)}\left(  \partial_{x}u_{x}%
+\partial_{y}u_{y}\right)  .
\end{equation}
This can be then used to simplify the in-plane components of the stress%
\begin{align}
T_{xx}  &  =\frac{E}{(1-\sigma^{2})}(\partial_{x}u_{x}+\sigma\partial_{y}%
u_{y}),\label{Txx}\\
T_{yy}  &  =\frac{E}{(1-\sigma^{2})}(\sigma\partial_{x}u_{x}+\partial_{y}%
u_{y}),\label{Tyy}\\
T_{yx}  &  =\frac{E}{2(1+\sigma)}(\partial_{x}u_{y}+\partial_{y}u_{x}).
\label{Txy}%
\end{align}
These expression are used to evaluate the first two terms in the integrand in
Eq.\ (\ref{Power_integral_ex}). The evaluation of the last term in the
integrand turns out to depend on whether we are looking at the in-plane or
flexural modes, and we now consider each case in turn.

\subsubsection{In-plane modes}

For the in-plane modes in the thin plate limit it is sufficiently
accurate to approximate $T_{zx}\simeq0$, and we can evaluate the
remaining
terms in $P$ with the approximations $u_{x}\simeq\bar{u}_{x}$, $u_{y}%
\simeq\bar{u}_{y}$ independent of $z$. This yields
\begin{equation}
P=\operatorname{Re}\left\{  -\frac{i\omega Ed}{4\left(  1-\sigma^{2}\right)
}\int dy\left[  2(\partial_{x}\bar{u}_{x}+\sigma\partial_{y}\bar{u}_{y}%
)\bar{u}_{x}^{\ast}+(1-\sigma)\left(  \partial_{x}\bar{u}_{y}+\partial
_{y}\bar{u}_{x}\right)  \bar{u}_{y}^{\ast}\right]  \right\}  .
\label{Poynting-inp}%
\end{equation}

\subsubsection{Flexural modes}

\label{Subsec_Power_Flexural}For the flexural mode the approximations
$T_{zx}\simeq0$ and $u_{z}(x,y,z)\simeq\bar{u}_{z}(x,y)$ independent of $z$
lead to the expressions for the horizontal displacements%
\begin{align}
u_{x}(x,y,z)  &  \simeq-z\bar{u}_{z}(x,y),\label{Flex_ux}\\
u_{y}(x,y,z)  &  \simeq-z\bar{u}_{z}(x,y). \label{Flex_uy}%
\end{align}
Using these expressions with Eqs.\ (\ref{Txx}-\ref{Tyy}) shows that the first
two terms in Eq.\ (\ref{Power_integral_ex}) are of order $d^{3}$, i.e.
\emph{third order} in the expansion parameter of thin plate theory $d/W$. It
turns out that to this order, we \emph{cannot} neglect the last term in
$T_{zx}$, even though all $z$-components in the stress tensor are nominally
``small''. Indeed comparing the group velocity evaluated from
Eq.\ (\ref{Power_integral_ex}) neglecting the term in $T_{zx}$ with those
given by numerically differentiating the dispersion curve shows a clear
discrepancy. This same problem comes up in deriving the wave equation for the
flexural waves%
\begin{equation}
\rho d\omega^{2}\bar{u}_{z}=D\nabla_{\perp}^{4}\bar{u}_{z}. \label{flex_wave}%
\end{equation}
The term on the left hand side is the mass per unit area times the vertical
acceleration, which is given by the integral over the depth of $\partial
_{x}T_{zx}+\partial_{y}T_{zy}$. Clearly the components of $T_{zi}$ cannot be
neglected completely. Their ``smallness'' is what leads to the unusual fourth
order derivative appearing in this wave equation, with a coefficient again
proportional to $d^{3}$.

We have used two methods to arrive at the correct calculation of the energy
flux integral for the flexural waves, which is then used to calculate the
group velocity for the these waves. The first is to use an improved
approximation to the expressions for the in-plane displacements Eqs.
(\ref{Txx},\ref{Tyy}) and a nonzero $T_{zx}$ following the approach of
Timoshenko \cite{T}. The second evaluates the energy flux in terms of the
vertical displacement and an effective vertical force, and in addition the
rotational displacement $\theta$ and corresponding torque $M$, as is used in
the macroscopic derivation\cite{LF7} of the wave equation (\ref{flex_wave}).
Either of these methods leads to the expression for the energy flux
\begin{multline}
P\simeq\frac{1}{2}\omega D\operatorname{Re}\left\{  \int dy\left[
2k^{3}\bar{u}_{z}\bar{u}_{z}^{\ast}+k(1-\sigma)(\partial_{y}\bar{u}_{z}%
)\partial_{y}\bar{u}_{z}^{\ast}-k(1+\sigma)(\partial_{y}^{2}\bar{u}_{z}%
)\bar{u}_{z}^{\ast}\right]  \right. \label{Poynting-power}\\
\left.  +Dk\left[  (1-\sigma)(\partial_{y}\bar{u}_{z})\bar{u}_{z}^{\ast
}\right]  _{y=\frac{W}{2}}-Dk\left[  (1-\sigma)(\partial_{y}\bar{u}_{z}%
)\bar{u}_{z}^{\ast}\right]  _{y=-\frac{W}{2}}\right\}
\end{multline}
The derivations are displayed in Appendix \ref{App_TPP}. The comparison of the
group velocity derived from Eq.\ (\ref{Poynting-power}) and from numerically
differentiating the dispersion curve now shows agreement to high accuracy.

\section{Scattering Analysis}

\label{Sec_scattering}The thermal attenuation is calculated from
Eq.\ (\ref{gamma-thin}) for normalized mode displacement fields or
(\ref{gamma_thin_nonorm}) in general. The group velocity for each mode can be
accurately evaluated numerically from the equality $v_{g}=2P/\rho w^{2}$, with
the energy flux $P$ given by Eq.\ (\ref{Poynting-inp}) for the in plane modes
and Eq.\ (\ref{Poynting-power}) for the flexural modes (both expressions are
for normalized displacement fields). These are all explicit results in terms
of the mode displacements, which are given by Eqs.\ (\ref{even-ux-filed}%
-\ref{odd-uy-field}) for the in-plane modes, and Eqs.\ (\ref{even_flex}%
,\ref{odd_flex}) for the flexural modes.

Before analyzing the scattering behavior, we first need to have a good
understanding of the dispersion relation of the modes, since the scattering
rates are strongly dependent on this.

\subsection{Dispersion relation and group velocity}

\begin{figure}[ptb]
\begin{center}
\includegraphics[ height=2.7198in, width=3.704in ]{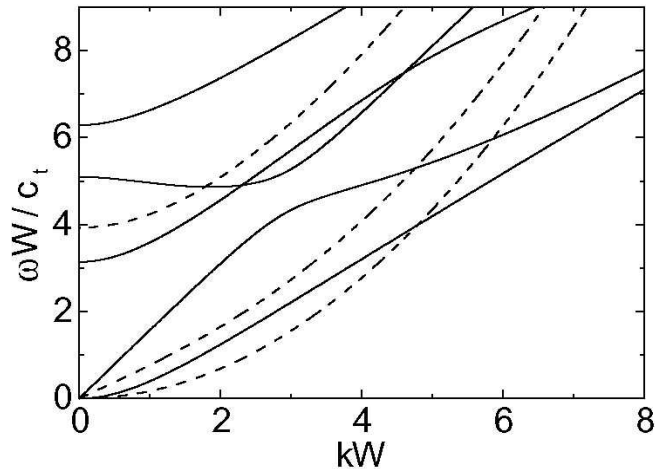}
\end{center}
\caption{Dispersion relation for in-plane modes (solid) and flexural modes
(dashed) for a geometry ratio $d/W=0.375$ and Poisson ration $0.24$. The wave
numbers are scaled with the width $W$, and the frequencies by $W/c_{t}$ with
$c_{t}=\sqrt{\mu/\rho\text{.}}$}%
\label{dispersion}%
\end{figure}

\begin{figure}[tbh]
\begin{center}
\includegraphics[ height=2.7717in, width=3.6832in ]{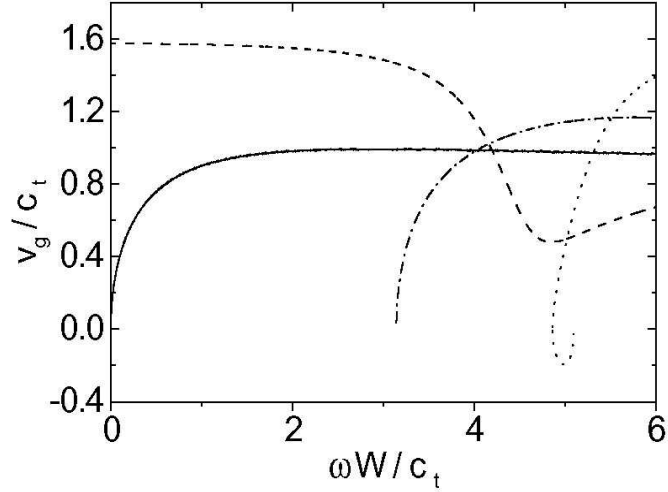}
\end{center}
\caption{Group velocity for in-plane modes for the same parameters as
Fig.\ (\ref{dispersion}): dash dotted - in-plane bending mode; solid -
compression mode. The wave numbers are scaled with the width $W$, and the
group velocities by $c_{t}$ with $c_{t}=\sqrt{\mu/\rho}$.}%
\label{inpvg}%
\end{figure}The dispersion relations for a representative case are shown in
Fig.\ (\ref{dispersion}). For this example we have used a Poisson
ratio of $0.24$, and a depth to width ratio of $d/W=0.375$, values
corresponding to the experimental work of Schwab et al.
\cite{SHWR00}. As we have discussed, the modes fall into four
classes, depending on their parity signatures. We label the lowest
mode from each class, the one with zero frequency as the wave
number goes to zero, as mode $0$, and the modes with successively
higher cutoff frequencies in each class as mode $1$, mode $2$,
etc., in that class.

Notice that one of the curves in the figure, the one for the
in-plane mode with cutoff frequency $\omega W/c_{t}\simeq5$, shows
anomalous dispersion with the frequency \emph{decreasing} as the
wave number increases up to about $3W^{-1}$. (This is actually an
even mode, and some higher even and odd modes also show anomalous
dispersion.) The dispersion curves for all modes $n>0$ have zero
slope, and so zero group velocity, at onset. As we will see later,
this results in a diverging scattering rate at each mode onset.
For the $n=0$ modes, as $\omega\rightarrow0$ two of the modes (the
compression and torsion modes) have linear dispersion, whilst the
other two lowest modes (in-plane and flexural bending modes)
exhibit quadratic dispersion. Figure (\ref{inpvg}) shows the group
velocities $v_{g}$ for the four lowest in-plane modes. The group
velocity of the bending mode approaches zero as
$\omega\rightarrow0$ whilst that of the compression mode becomes
constant. The group velocity of the compression mode suddenly
drops to $\sim0.5c_{t}$ around $\omega W/c_{t}\sim4.6,$ then
gradually recovers and approaches $0.9c_{t}$. These features of
the dispersion curve will be reflected in the behavior of the
scattering of the waves.

\subsection{Scattering behavior}

We first consider the scattering and reduction of the thermal transport by
white noise roughness $\tilde{g}(k)=\tilde{g}(0)$. This allows us to focus on
the role of geometry and the unusual mode structure of the elastic waves in
the physics.

\squeezetable
\begin{table}[ptb]
\begin{ruledtabular}
\begin{tabular}
[c]{cccccc} & $\omega/\sqrt{E/\rho}$ & $v_{g}/\sqrt{E/\rho}$ &
$\phi_{x}$ & $\phi_{y}$ & $\phi_{z}$\\\hline
Extension & $k$ & $1$ & $1$ & $O(ky)$ & $O(kz)$\\
In-plane bend & $(w/\sqrt{12})k^{2}$ & $(w/\sqrt{3})k$ & $-iky$ &
$1$ &
$O(kz)$\\
Torsion & $\sqrt{2/(1+\sigma)}(d/w)k$ & $\sqrt{2/(1+\sigma)}(d/w)$
& $O(kyz)$
& $-z$ & $y$\\
Flex-bend & $(d/\sqrt{12})k^{2}$ & $(d/\sqrt{3})k$ & $-ikz$ &
$O(k^{2}yz)$ &
$1$%
\end{tabular}
\end{ruledtabular}
\caption{Dispersion relation, group velocity, and (unnormalized) transverse mode
structure for the four modes with zero frequency at zero wave vector.}%
\label{Table_Modes}%
\end{table}

In the low frequency limit the dispersion curve and the spatial dependence of
the modes take on the simple analytic forms shown in Table \ref{Table_Modes},
allowing us to make analytic predictions for the scattering at low
frequencies, and then the thermal conductance at low temperatures. Since only
small wave vector scattering is involved in these calculations, the results
are true for a general roughness correlation function, providing $\tilde
{g}(0)$ is nonzero. The mode structure in Table \ref{Table_Modes} may be
calculated from Eqs. (\ref{even-ux-filed}-\ref{odd_flex}) taking
$k\rightarrow0$ or from arguments of macroscopic elasticity theory.

\squeezetable
\begin{table}[ptb]
\begin{ruledtabular}
\begin{tabular}
[c]{c|c}
in-plane & flexural\\\hline%
\begin{tabular}
[c]{c|c|c}%
cc & bb & bc,cb\\\hline
$2\bar{\omega}^{2}$ & $\sqrt{3}\bar{\omega}$ & $\frac{3^{5/4}}{2^{3/2}}%
\bar{\omega}^{3/2}$%
\end{tabular}
&
\begin{tabular}
[c]{c|c|c}%
tt & bb & tb,bt\\\hline $\frac{9(1+\sigma)}{4}\left(
\frac{W\bar{\omega}}{d}\right)  ^{2}$ & O$\left[  \left(
\frac{W\bar{\omega}}{d}\right)  ^{3}\right]  $ &
$\frac{3^{5/4}(1+\sigma)^{1/2}}{4}\left(
\frac{W\bar{\omega}}{d}\right)
^{3/2}$%
\end{tabular}
\end{tabular}
\end{ruledtabular}
\caption{Scattering coefficients for the zero onset frequency modes at low
frequencies: c denotes compression, b denotes bend, t denotes
torsion, bb
denotes bend to bend scattering etc. Values are quoted for $\gamma_{m}W^{4}%
/\tilde{g}(0)$ as a function of scaled frequency
$\bar{\omega}=\omega c_{E}/W$. For the flexural bend to bend
scattering (bb) the terms in the braces in Eq.\ (\ref{gamma-thin})
cancel to leading order resulting in very small
$O(\bar{\omega}^{3})$ scattering. There is no scattering between
in-plane and flexural modes for the z-independent roughness assumed.}%
\label{Table_Scattering}%
\end{table}

\begin{figure}
[ptb]
\begin{center}
\includegraphics[
height=3in,
width=3.9998in
]%
{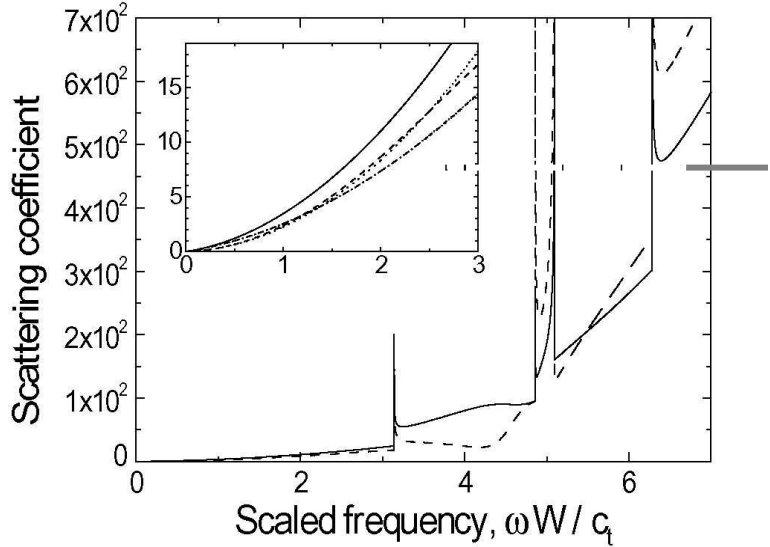}%
\caption{Attenuation coefficient $\gamma_{m}W^{4}/\tilde{g}(0)$ for scattering
from the two lowest $m=0$ inplane modes to any other mode as a function of
scaled frequency $\omega W/c_{t}$: solid line - inplane bend mode; dashed line
- compression mode. The insert shows an enlargement of the low frequency
region, and compares with the analytic low frequency expressions from Table
\ref{Table_Scattering}: dotted - analytic inplane bend mode; dash-dotted -
analytic compression mode; other lines as in the main figure.}%
\label{Fig_scattinp0}%
\end{center}
\end{figure}
The contributions to the thermal attenuation coefficient in the
low frequency limit ($\omega W/c_{t}\ll1$) from the various
scattering processes are shown in the Table \ref{Table_Scattering}
\footnote{A more accurate expression for the scattering between
the in-plane compression and bending modes is giving by keeping
the next order term in $\bar{\omega}$ which is
$O(\bar{\omega}^{1/2})$ for this scattering process. To include
this correction multiply the expression in the table by
$(1+\sqrt{\bar{\omega}}/12^{1/4})^{2}$.}. The expressions take on
their simplest form if we introduce the frequency scaled with the
velocity of the long wavelength compression mode
$\bar{\omega}=\omega c_{E}/W$ with
$c_{E}=\sqrt{E/\rho}=\sqrt{2(1+\sigma)}c_{t}$. The power laws can
largely be understood from the prefactor in
Eq.~(\ref{gamma_thin_nonorm}), $\gamma
_{n,m}\propto\omega^{2}/v_{g}^{\left(  m\right) }v_{g}^{\left(
n\right) }$. The group velocity $v_{g}$ becomes a constant at
small frequencies for the compression and torsion modes. Thus the
torsion-torsion and compression-compression scattering shows the
$\omega^{2}$ dependence corresponding to Rayleigh scattering in
one dimension, and as was found for scalar waves with linear
dispersion. On the other hand for the bending modes
$v_{g}\propto\omega^{1/2}$. This has the important consequence
that the in-plane bend-bend scattering increases more rapidly at
low frequencies proportional to $\omega$, and the torsion-bend and
compression-bend scattering have an $\omega^{3/2}$ frequency
dependence. For the flexural bend-bend scattering the two terms in
the braces in Eq.\ (\ref{gamma-thin}) cancel to leading order
resulting in smaller scattering $O(\omega^{3})$ than given by the
prefactor alone. Note that the expressions for the flexural modes
involve additional factors of $W/d$, so that these modes will be
scattered more strongly at a given $\omega$ in the thin plate
limit. This is because these
modes are softer, so that the scattering wave vectors are larger for the same frequency.%

\begin{figure}
[ptb]
\begin{center}
\includegraphics[
height=3in, width=3.9998in
]%
{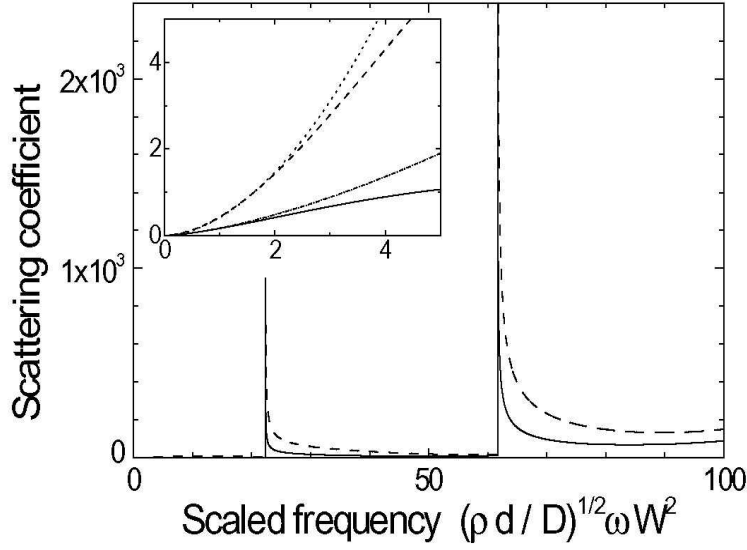}%
\caption{Attenuation coefficient $\gamma_{m}W^{4}/\tilde{g}(0)$ for scattering
from the two lowest $m=0$ flex modes to any other mode as a function of scaled
frequency $\omega\sqrt{12(1-\sigma^{2})}(W/d)W/c_{E}$: solid line - the
flex-bend mode; dashed line - torsion mode. The insert shows an enlargement of
the low frequency region, and compares with the analytic low frequency
expressions from Table \ref{Table_Scattering}: dotted line - analytic
approximation for the flex-bend mode; dash-dotted: analytic expression for the
torsion mode; other lines as in the main figure.}%
\label{Fig_scatflex0}%
\end{center}
\end{figure}

Numerical results for the attenuation coefficient $\gamma_{m}$ of
the four lowest modes are shown in Fig.\ (\ref{Fig_scattinp0}) for
the in-plane and Fig.\ (\ref{Fig_scatflex0}) for the flexural
modes. The plot for the in-plane modes in particular shows
interesting structure deriving from the complicated dispersion
curves of Fig.\ \ref{dispersion}. Much of this structure can be
understood from the product of group velocities in the denominator
of Eq.\ (\ref{gamma-thin}). In particular there is a square root
divergence in $\gamma_{m}$ at the onset frequency of each mode
where the group velocity is zero. In addition, the large
scattering around $\omega W/c_{t}=5$ derives from the region of
anomalous dispersion, since the group velocity is small in this
frequency range. The insert to Fig.\ (\ref{Fig_scattinp0})\ shows
an expanded view of the low frequency behavior, using the results
from Table \ref{Table_Scattering} together with the next order
correction for the compression-bend scattering. The agreement for
the compression mode is very good even up to $\omega
W/c_{t}\sim3$, whereas for the bend mode the correspondence is
only good for $\omega W/c_{t}\lesssim0.5$. The scattering for the
flexural modes shows generally similar results, Fig.\
(\ref{Fig_scatflex0}) although the behavior is simpler
corresponding to the rather featureless dispersion curves. At low
frequencies, (insert to Fig.\ (\ref{Fig_scatflex0})), the
scattering of the flexural-bend mode is
small, since the intramode scattering is reduced by the cancellation discussed above.%

\begin{figure}
[ptb]
\begin{center}
\includegraphics[
height=3in, width=3.9998in
]%
{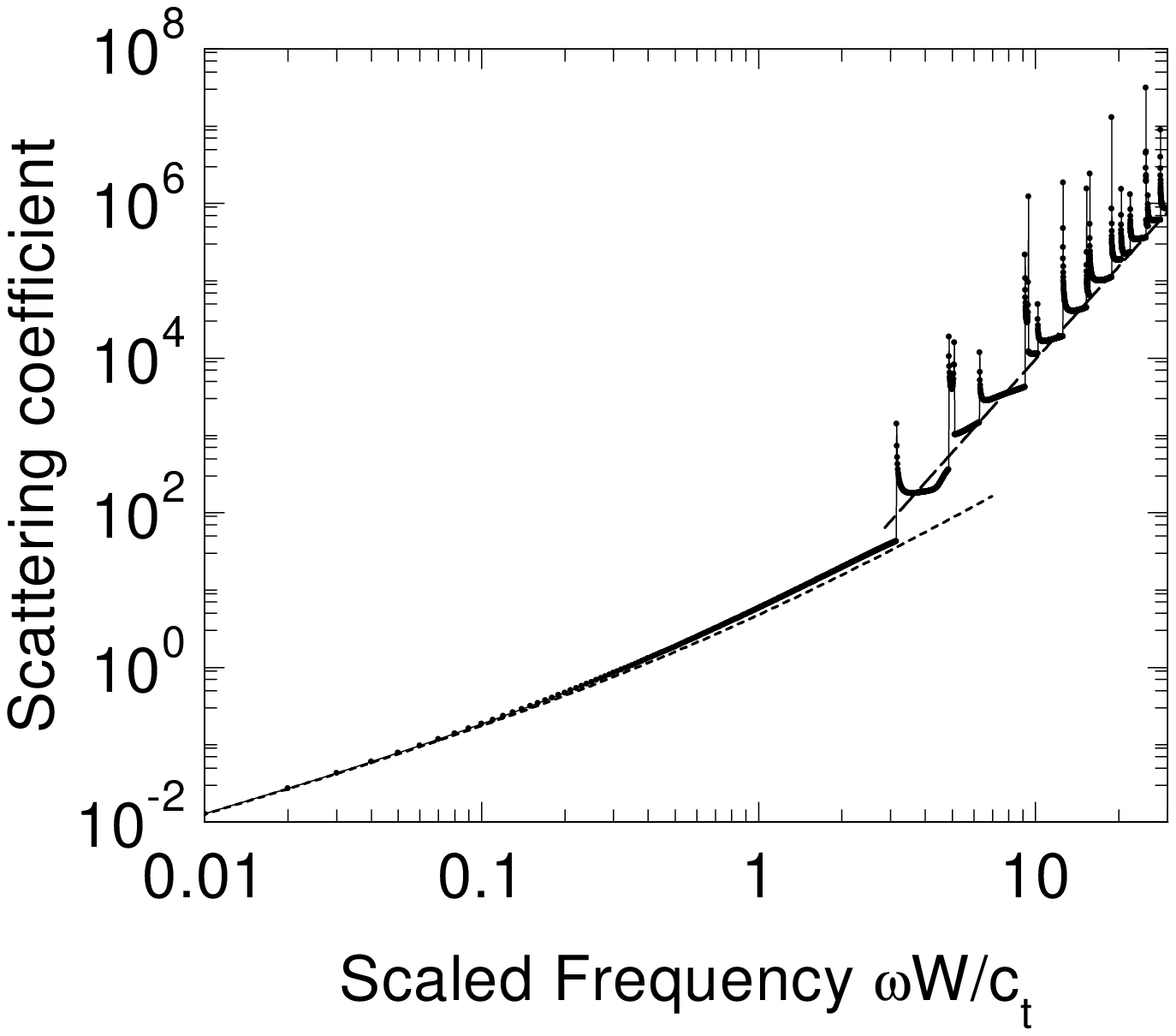}%
\caption{Total scattering $\sum_{m}\gamma_{m}W^{4}/\tilde{g}(0)$ for the
in-plane modes on a log-log plot. The dotted line shows the low frequency
analytic expression from Table \ref{Table_Scattering}, and the dashed line
shows a power law $4$. (Note that the heights of the peaks in the plot are not
significant, depending on how close the individual points, separated by $0.01$
in $\omega W/c_{t}$, used in constructing the plot are to the mode onset
frequencies where the scattering diverges.)}%
\label{scatt_loglog}%
\end{center}
\end{figure}


Figure\ (\ref{scatt_loglog}) shows the total scattering
$\sum_{m}\gamma_{m}$ for the in-plane modes on a log-log plot,
again with white noise roughness. At very low frequencies the
scattering varies proportional to $\omega$ corresponding to the
dominant intramode scattering of the compression mode at low
frequencies (Table \ref{Table_Scattering}). For frequencies up to
$\omega W/c_{t}\simeq3.5$, the first nonzero onset frequency of an
in-plane mode, the analytic low-frequency expression given by
summing the in-scattering expressions from Table
\ref{Table_Scattering} (cc, cb, bc, and bb), shown as the dotted
line in Fig.\ \ref{scatt_loglog}, gives a good approximation to
the full results. At higher frequencies the total scattering
increases rapidly, following a general trend proportional to
$\omega^{4}$ (dashed line) together with divergent scattering at
each mode onset frequency. The $\omega^{4}$ power law can be
understood as the combination of the explicit $\omega^{2}$
dependence of Eq. (\ref{gamma-thin}), together with two powers of
$\omega$ coming from the number of modes available for scattering
from and to.


\subsection{Change in the thermal conductance}

In the weak scattering limit the change in thermal conductance at low
temperatures can be derived directly from the expressions for the scattering
at low frequencies. If we write the thermal attenuation coefficient of mode
$m$ as $\gamma_{m}L=A(\omega/\omega_{0})^{p}$, where $p$ is the power law
obtained in the low frequency limit and $\omega_{0}$ some characteristic
frequency, then the corresponding contribution of the suppression of the
thermal conductance from this mode is
\begin{equation}
\delta K_{m}/K_{u}=AI_{p}(T/T_{0})^{p} \label{temp-power}%
\end{equation}
with $T_{0}=\hbar\omega_{0}/k_{B}$ \ the corresponding characteristic
temperature and $K_{u}=\pi^{2}k_{B}^{2}T/3h$ the universal thermal
conductance. The constant $I_{p}$ can be obtained evaluating the integral%
\begin{equation}
I_{p}=\frac{3}{\pi^{2}}\int_{0}^{\infty}dy\frac{y^{p+2}e^{y}}{(e^{y}-1)^{2}}.
\end{equation}
Thus the power law for the temperature dependence of the depression of the
thermal conductivity is the same as the one for the low frequency behavior of
the scattering coefficient.

\begin{figure}
[ptb]
\begin{center}
\includegraphics[
height=3in, width=3.9998in
]%
{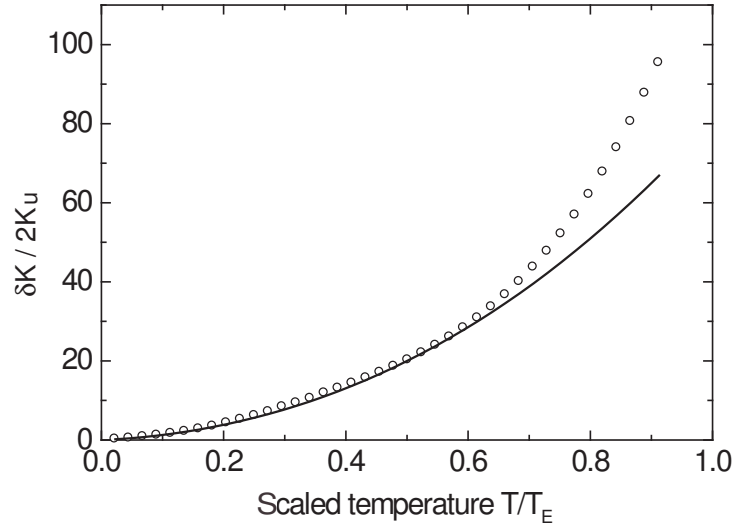}%
\caption{Reduction in the thermal conductance scaled with the universal
conductance $K_{u}$ for the lowest in-plane modes as a function of
scaled temperature $T/T_{E}$ with $T_{E}=\hbar c_{E}/k_{B}W$:
solid line - low temperature analytical expressions from Table
\ref{Table_Scattering}: points - full expression evaluated
numerically. The quantity plotted is $(\delta K_{c}+\delta
K_{ib})/2K_{u}$ with $\delta K_{c},\delta K_{ib}$ the depression
of the contributions to the conductance by the scattering for the
compression
and in-plane bending modes.}%
\label{dK-inplane}%
\end{center}
\end{figure}

\begin{figure}
[ptbptb]
\begin{center}
\includegraphics[
height=3in, width=3.9998in
]%
{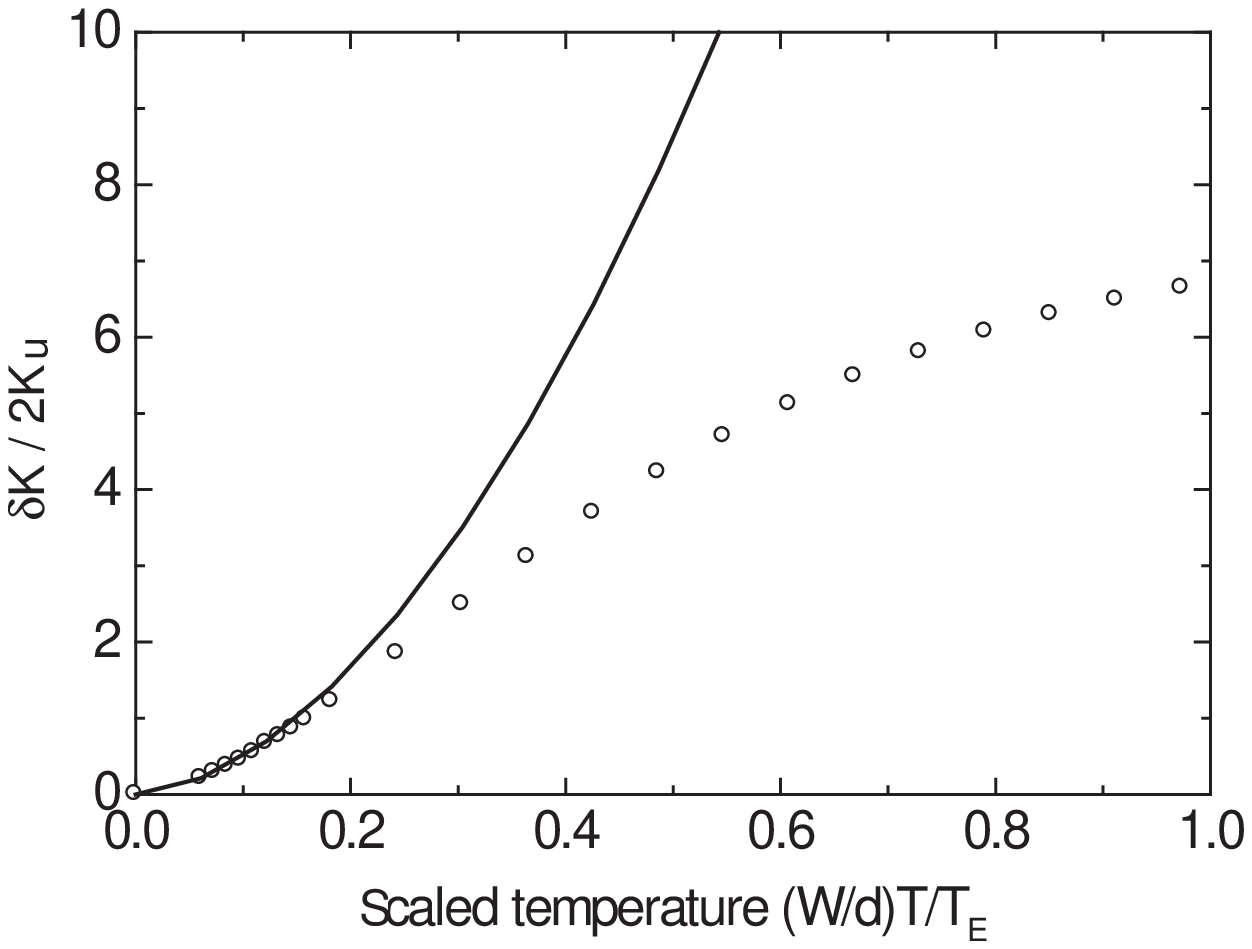}%
\caption{Similar to Fig.$\;$(\ref{dK-inplane}), $\delta K/2K_{u}$ for the
lowest flexural modes (torsion and flexural-bending) as a function of the
scaled temperature $T/T_{F}$ with $T_{F}=\hbar c_{E}d/k_{B}W^{2}$.}%
\label{dK-flex}%
\end{center}
\end{figure}

Figures (\ref{dK-inplane}) and (\ref{dK-flex}) show the thermal conductance
depression scaled with the universal value $K_{u}$ as a function of the
appropriate scaled temperature for the lowest in-plane and flexural modes,
showing the deviation from the low temperature power laws as the temperature
is raised. For the in plane modes we use the characteristic temperature
$T_{E}=\hbar c_{E}/k_{B}W$ and for the flexural modes $T_{F}=\hbar
c_{E}d/k_{B}W^{2}$. The individual plots are then independent of the geometry.
To combine the contributions from the in-plane and flexural modes the ratio
$d/W$ is needed to relate the two temperature scale factors. In the thin plate
limit $T_{F}=(d/W)T_{E}\ll T_{E}$.%

\section{Comparison with Experiment}

\label{Sec_experiment}

\subsection{Experimental geometry}

Based on the SEM micrograph of the experimental
structure\cite{Ssp}, we set the dimensions of the structure in the
following way. In the experimental structure of Schwab et al.\ the
thermal pathway was constructed with the shape function
$W(x)=W\cosh(Ax)$ so that the beam width becomes large and joins
smoothly to the thermal reservoirs at the ends, reducing the
scattering due to the geometric imperfection at these junctions.
Unfortunately this makes the calculation of the behavior of the
elastic waves in the beams much harder. However both with and
without the scattering off surface roughness, we expect the narrow
portion of the beam to dominate the behavior. Thus we simplify the
structure and model it as an elastic beam with rectangular cross
section of width $W$, depth $d$, and effective length $L$. We
estimate the width as the narrowest width of the structure,
$W\simeq160nm$, and $L=1$ $\mu$m as the length over which the
width is approximately constant. The thickness of the material was
$d=60$ nm. The accuracy of the length estimation is not very
critical, since the only length dependence in the scattering rate
$\gamma$ appears in the combination $\delta^{2}L$ where $\delta$
is the rms roughness which is a parameter of the model, so that
any error in the assignment of $L$ will just change the value
assigned to $\delta $. The width $W$ on the other hand plays a
crucial role, for example determining the frequency cutoffs of the
various modes, and so the temperature dependence of the thermal
conductivity.

\subsection{Roughness correlation function}

Since the nature of the surface roughness on the experimental structure is not
known, to fit the experimental data we need a sensible parameterization of the
roughness. As a starting point we choose a Gaussian correlation function for
the roughness, leading to the spectral density%
\begin{equation}
\tilde{g}\left(  k\right)  =\sqrt{\pi}a\delta^{2}\exp\left[  -\frac{a^{2}}%
{4}k^{2}\right]  . \label{Gaussian}%
\end{equation}
This parameterization of the roughness contains two parameters: $\delta$ the
rms roughness and $a$ the correlation length.

To analyze the data, we first quantify the amount of scattering by
subtracting the data of Schwab et al. from the ideal thermal
conductance obtained numerically using the ``xyz''
algorithm\cite{NAW97}. Then we attempt to fit the data by
adjusting the two parameters $a$ and $\delta^{2}L$.
\begin{figure}
[ptb]
\begin{center}
\includegraphics[
height=2.5486in,
width=3.8233in
]%
{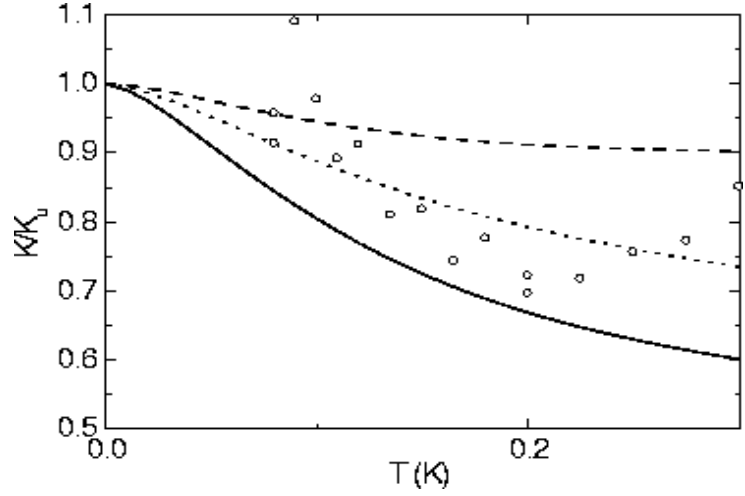}%
\caption{Attempts to fit the low temperature data $T\lesssim0.2$ K using
various values of $a\delta^{2}$: solid line - $\sqrt{\pi}a\delta^{2}=0.1$;
dotted line - $\sqrt{\pi}a\delta^{2}=0.05$; dashed line - $\sqrt{\pi}%
a\delta^{2}=0.02$; open circles - from the experimental data of Schwab et al.}%
\label{a-delta2}%
\end{center}
\end{figure}

The inadequacy of Eq.\ (\ref{Gaussian}) in fitting the experimental data is
shown by the low temperature fits in Fig.\ (\ref{a-delta2}). At these low
temperatures only small wave number modes are excited, so that the exponential
term in Eq.\ (\ref{Gaussian}) can be approximated as unity and $\tilde
{g}\left(  k\right)  \simeq\tilde{g}\left(  0\right)  =\sqrt{\pi}a\delta^{2}$.
Thus the roughness parameters only appear in the combination $a\delta^{2}$,
and this quantity can be varied to attempt to fit the low temperature region.
As seen from the figure, increasing $a\delta^{2}$ causes scattering that is
systematically larger than the experimental data at the low temperatures,
while decreasing $a\delta^{2}$ does not provide enough scattering in the range
$0.1<T<0.2$ K.

Although there is considerable scatter in the data over the range of the fit,
the systematic differences between the predictions and the data lead us to
propose a modified form of the roughness correlation that reduces the
scattering at small wave numbers%
\begin{equation}
\tilde{g}\left(  k\right)  =\sqrt{\pi}a\delta^{2}\exp\left[  -\frac{a^{2}%
\left(  k-k_{0}\right)  ^{2}}{4}\right]  . \label{Gaussian-shift}%
\end{equation}
A nonzero value of the parameter $k_{0}$ leads to a roughness
correlation function that is maximum at a length scale of order
$k_{0}^{-1}$, and serves to reduce the scattering at long
wavelengths. As mentioned in the introduction section, the same
discrepancy (i.e., the overestimation of the scattering at long
wavelengths in the theory compared with experiment) was found
using the scalar model of the elastic waves\cite{SC00}. The full
elasticity theory considered here actually makes the discrepancy
worse, since the scattering at small frequencies now is predicted
to increase more rapidly at small frequencies than the
$\omega^{2}$ found in the scalar theory, varying as $\omega^{p}$
with
$p<2$ for most of the scattering processes, see Table \ref{Table_Scattering}.%

\begin{figure}
[ptb]
\begin{center}
\includegraphics[
height=2.4751in,
width=3.3252in
]%
{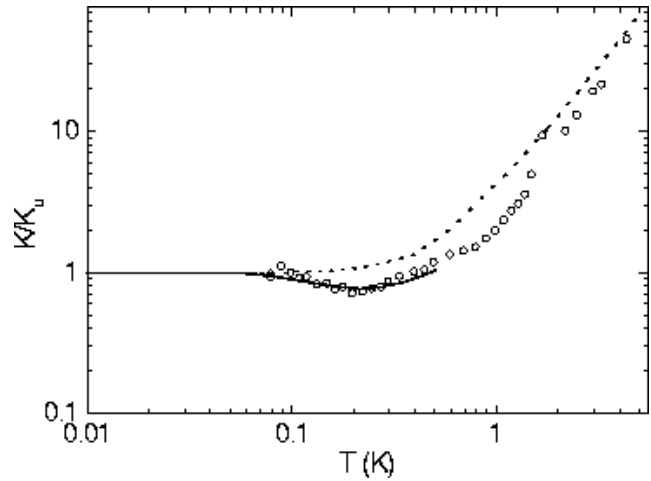}%
\caption{Thermal conductance per mode scaled with universal value $K_{u}$:
solid line - fit using roughness parameters $a/W=5.5$, $\delta/W=0.2$, and
$k_{0}W=4.9$; circles - data of Schwab et al. The dotted line shows the ideal
value with no scattering.}%
\label{fitK}%
\end{center}
\end{figure}
To fit the data of Schwab et al, we need to determine three
parameters: $k_{0}$, $a$, and $\delta$. As well as looking at the
fit by eye, we evaluate the quality of the fit by calculating the
mean square deviation of the data from the theory curve over the
temperature range up to $0.4K$. At higher temperatures many modes
becoming excited, and the scattering of individual modes becomes
strong, so that our theory is less reliable. Since the onset
frequency of the scattering at low frequencies and the initial
decrease in thermal conductance with increasing temperature near
the onset is mainly determined by $k_{0}$, this parameter is the
easiest to determine. We find the value $k_{0}W=4.9$ , rather
insensitive to the values of $a$ and $\delta$. We have also
investigated the fit with a delta function noise correlation
function (i.e. $a\rightarrow0$ in Eq.\ (\ref{Gaussian-shift})).
With this roughness, to get a maximum in the decrease in the
thermal conductance at about $T\sim0.2K$ requires $k_{0}W$ values
between $4.5$ and $5$, which agrees with the previous estimate.%

\begin{figure}
[ptb]
\begin{center}
\includegraphics[
height=2.8253in,
width=3.614in
]%
{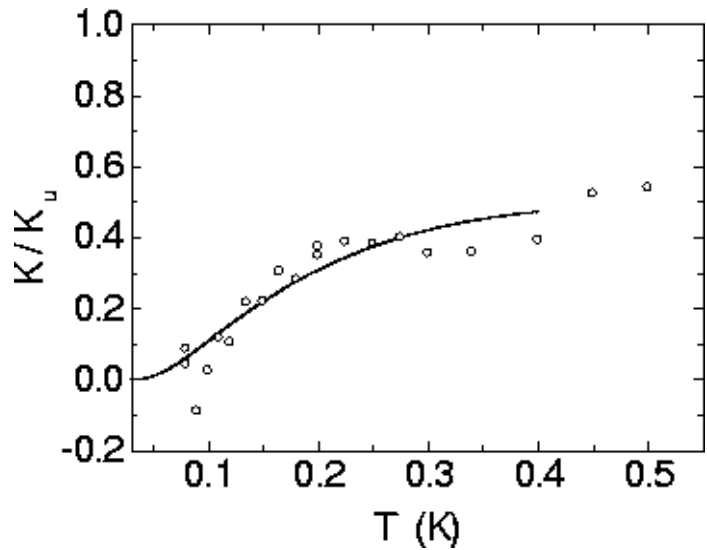}%
\caption{Same as in Fig.\ (\ref{fitK}) but showing the decrease of $K/K_{u}$
from the ideal value.}%
\label{fitdeltaK}%
\end{center}
\end{figure}
The remaining fit parameters $a$ and $\delta$\ are not well
determined. Visual inspection of the fit and a plot of the error
as a function of $(a,\delta)$ shows a reasonable fit over the
range $a/W=5,\delta/W=0.17$ to $a/W=6.7,\delta /W=0.32$. A choice
of the roughness amplitude towards the lower end of this range
seems most likely physically (e.g., $\delta/W=0.31\simeq50$ nm is
unlikely for roughness generated by chemical etch). We therefore
use $\delta/W=0.2$ and $a/W=5.5$ to generate the curves.

A difficulty of fitting the data is the lack of data points at
very low temperatures: it is in this range where only a few modes
are involved that we have a very good understanding of the
scattering. At higher temperatures many more modes become
involved, and the scattering of individual modes becomes strong,
so that the second order approximation used in calculating the
scattering will not be good. A\ full test of the theory explaining
the reduction in the thermal conductance in terms of the
scattering off surface roughness requires more data below a
temperature of about $0.08K$ for the type of geometry used by
Schwab et al., or systems with smaller geometries where the
effects can be measured at higher temperatures.

\subsection{Individual mode contribution to the thermal conductance}%

\begin{figure}
[ptb]
\begin{center}
\includegraphics[
height=2.5763in,
width=3.8649in
]%
{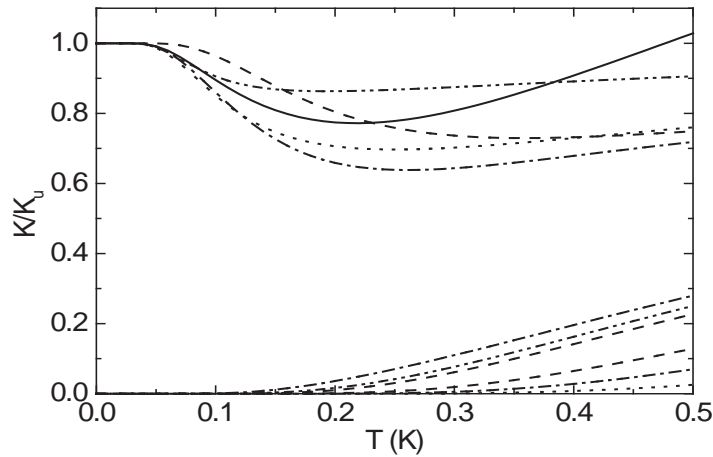}%
\caption{Individual mode contribution to the thermal conductance. The lowest
two flex modes and lowest three in plane modes are shown. The
contributions to $K/K_{u}$ from the four modes with zero onset
frequency tend to unity at low temperatures. The higher modes only
contribute at higher temperature. The modes are: dash-dotted -
in-plane bending; dashed - compression; dotted - torsion;
dashed-dotted-dot - out-of plane bending. The solid line shows the
sum of all the mode contributions, reduced by $4K_{u}$. Values of
the roughness parameters used were $a/W=5.5$, $\delta/W=0.2$,
$k_{0}W=4.9$, and
$d/W=0.375$.}%
\label{Kmode}%
\end{center}
\end{figure}

It is interesting to investigate the contribution to the total
thermal conductance of the individual modes with the roughness
parameters used to fit the experimental data. This is shown in
Fig.\ (\ref{Kmode}). The flex-bending mode shows a much smaller
contribution to the reduction in $K$ at low temperatures for the
reason we have already discussed. The modes with nonzero onset
frequencies start to contribute significantly above about
$T\simeq0.2K$, and this is the predominant cause for the increase
in thermal conductivity above this temperature, since the recovery
of the thermal conductance for the lowest mode occurs very slowly.

\section{Conclusion}

\label{Sec_conclusion}We have investigated the effect of surface roughness on
the scattering of elastic waves in a rectangular beam or
waveguide, and the resulting depression of the thermal conductance
in the low temperature quantized limit, using full elasticity
theory. Our formulation is quite general, but to obtain concrete
results we have specialized to the thin-plate limit, which should
be a reasonable approximation for many mesoscopic experiments
where the depth of the structures is fixed by the epitaxial
growth, whilst the width is determined lithographically. The thin
plate limit preserves the peculiar features of the elastic waves
in the full elastic theory, namely a quadratic dispersion at long
wavelengths for two of the low frequency modes, and regions of
negative dispersion in the spectra. A robust result is that the
low frequency asymptotic dependence of the scattering by
unstructured roughness of the modes that propagate at low
frequencies (the ones that are important in the low temperature
universal thermal conductance) depends on the structure of the
modes and the dispersion relation, and is \emph{not} the simple
$\omega^{2}$ dependence of Rayleigh scattering as found in the
scalar approximation to the modes. We find different power laws
for the various mode scattering processes that can be understood
largely from the dispersion relations at: $\omega$ for intramode
scattering for the in-plane bend mode (the flex-bend intramode
scattering is anomalous because of a cancellation between leading
order terms, and varies as $\omega^{3}$); $\omega^{3/2}$ for
scattering between the bend modes and the modes with linear
dispersion (torsion and compression modes); and the usual
$\omega^{2}$ for the intramode scattering of the modes with linear
dispersion. The current experimental data on the suppression of
the low temperature thermal conductance below the universal value
does not extend to low enough temperatures to provide a good test
of these predictions. To investigate this prediction further, it
would be interesting to extend the experiments to lower
temperatures, or to smaller devices such as carbon nanotubes,
where the characteristic temperature scales (when a typical
thermally excited phonon has a wavelength comparable with the
device dimensions) are higher.

We have used our results to understand the data the data of Schwab
et al.,\ who observed a depression of the thermal conductance
below the universal value in the temperature range of $0.1K$ to
$0.4K$. Although the scatter in the data is considerable at these
low temperatures, the observations seem to show a delay in the
onset of the depression scattering as the temperature is raised,
beyond what can be fitted with our predictions for unstructured
surface roughness. We tentatively resolve this delay by supposing
that the surface roughness has a maximum amplitude at some nonzero
length scale, which we parameterize by a shifted Gaussian
correlation function. Due to the lack of data at low temperatures,
a precise determination of the roughness parameters is not
possible. However, we do obtain a fit to the data with parameters
that do not look unreasonable when compared with electron
micrographs of the actual devices.

Our results are based on second order perturbation theory, and the
thermal conductance is evaluated assuming the scattering over the
length of the device is small. This is a good approximation at low
temperatures, but the scattering becomes strong at higher
temperatures, particularly for the new modes excited as the
temperature is raised, which have a diverging scattering at onset
due to the flat dispersion relation here. At higher temperatures
multiple scattering and perhaps phonon localization will therefore
become important. Kambili et al.\cite{KFFL99} and Sanchez-Gil et
al. \cite{SFMY99}\ have numerically investigated the these effects
in the simplified scalar wave approximation. It would be
interesting in the future to extend their work to the full
elasticity model.

\begin{acknowledgments}
This work was supported by NSF grant no. DMR-9873573. We thank Ruben
Krasnopolsky for help on the numerical codes.
\end{acknowledgments}

\appendix

\section{Incident and Scattered Fields}

\label{Appendix_Separate}Using Green's theorem we have expressed the
displacement field at frequency $\omega$ in terms of the surface integral%
\begin{equation}
u_{q}(\mathbf{x})=\int_{S^{\prime}}\left[  n_{j}^{\prime}T_{ij}\left(
\mathbf{x}^{\prime}\right)  G_{iq}\left(  \mathbf{x}^{\prime},\mathbf{x}%
\right)  -n_{j}^{\prime}u_{i}\left(  \mathbf{x}^{\prime}\right)  \Gamma
_{ijq}\left(  \mathbf{x}^{\prime},\mathbf{x}\right)  \right]  dS^{\prime}
\label{total field2}%
\end{equation}
Eq.\ (\ref{total field2}) involves the integration over a closed surface
$S^{\prime}$, which we have chosen to be the smooth boundaries together with
the cross sections at $x^{\prime}\rightarrow\pm\infty$. In this appendix we
show that the integration over the sections at $\pm\infty$ simply yields the
incident field $u_{q}^{\mathrm{in}}$, and this allows us to deduce the
expression for the scattered field as an integration over the side surfaces.
To deduce this result, we first need to derive what are known as reciprocity
relations for the elastic modes\cite{A}.

Let $\mathbf{u}^{(r)}$ and $\mathbf{u}^{(s)}$ be the displacement fields for
modes $r$ and $s$ in the ideal beam, and $\mathbf{T}^{(r)}$, $\mathbf{T}%
^{(s)}$ the corresponding stress tensor fields. The modes satisfy the wave
equation at frequency $\omega$, so that%
\begin{equation}%
\begin{array}
[c]{c}%
\rho\omega^{2}u_{i}^{(r)}+\partial_{j}T_{ij}^{(r)}=0\\
\rho\omega^{2}u_{i}^{(s)}+\partial_{j}T_{ij}^{(s)}=0
\end{array}
\end{equation}
Multiply the first equation by $u_{i}^{(s)\ast}$\ and the complex conjugate of
the second by $u_{i}^{(r)}$, subtract the two equations, integrate over a
volume of the beam between $x=x_{1}$ and $x=x_{2}$, and finally use the
divergence theorem to find%
\begin{equation}
\int_{S}\left[  u_{i}^{(s)\ast}T_{ij}^{(r)}-u_{i}^{(r)}T_{ij}^{(s)\ast
}\right]  \hat{n}_{j}dS=0, \label{uttu}%
\end{equation}
where the integral is over the surface bounding the volume, consisting of the
sides of the beam between $x_{1}$ and $x_{2}$, and the sections at $x_{1}$ and
$x_{2}$. The integrations over the sides of the beam are zero by the stress
free boundary conditions. For the integration over the sections introduce the
explicit $x$-dependence $\mathbf{u}^{(r)}=\mathbf{\phi}(y,z)e^{ik_{r}x}$ and
$\mathbf{T}^{(r)}=\mathbf{\bar{T}}^{(r)}(y,z)e^{ik_{r}x}$ with $k_{r}$ the
wave number of mode $r$ at frequency $\omega$ etc. Then Eq.\ (\ref{uttu})
reduces to
\begin{equation}
\left(  1-e^{i\left(  k_{r}-k_{s}\right)  \left(  x_{1}-x_{2}\right)
}\right)  \int\int\left[  \phi_{i}^{(s)\ast}\bar{T}_{ix}^{(r)}-\phi_{i}%
^{(r)}\bar{T}_{ix}^{(s)\ast}\right]  dydz=0
\end{equation}
and the integral is independent of $x$. Unless the prefactor is zero, this
shows us that the integral over the section must be zero, and so
\begin{equation}
\int\int\left[  u_{i}^{(s)\ast}T_{ix}^{(r)}-u_{i}^{(r)}T_{ix}^{(s)\ast
}\right]  \,dydz=0,\quad k_{r}\neq k_{s}. \label{Eq_to-reciprocity}%
\end{equation}
This is one version of the reciprocity relations.

For our purposes it is more convenient to express the condition for the
reciprocity integral to be zero in terms of the group velocity rather than the
wave number. To do so, we need to consider the dispersion curves. The
condition for the reciprocity integral to be nonzero, $k_{r}=k_{s}$ for modes
$r,s$ at the same frequency $\omega$, actually implies $r$ and $s$ are the
\emph{same} mode, so that in fact $v_{g}^{(r)}=v_{g}^{(s)}$. The only other
possibility is that $r$ and $s$ are modes with dispersion curves that cross at
frequency $\omega$, $k=k_{r}=k_{s}$. However only modes of different $y,z$
parity signatures can cross, and then the integration over the section for
these different modes in Eq.\ (\ref{Eq_to-reciprocity}) is again zero. Thus we
can rewrite the reciprocity relation as%
\begin{equation}
\int\int\left[  u_{i}^{(s)\ast}T_{ix}^{(r)}-u_{i}^{(r)}T_{ix}^{(s)\ast
}\right]  \,dydz=0,\quad v_{g}^{(r)}\neq v_{g}^{(s)}. \label{Eq_reciprocity_1}%
\end{equation}
If $r$ and $s$ are the same mode, the integral is related to the energy flux
and hence to the group velocity (see Eq.\ (\ref{power}))
\begin{equation}
\int\int dydz\left(  u_{i}^{(r)\ast}T_{ij}^{(r)}-u_{i}^{(r)}T_{ij}^{(r)\ast
}\right)  =2i\rho\omega v_{g}^{(r)}. \label{Eq_reciprocity_2}%
\end{equation}

We now use Eqs.\ (\ref{Eq_reciprocity_1},\ref{Eq_reciprocity_2}) to evaluate
the contributions to Eq.\ (\ref{total field2}) from the integrations over the
sections at $x^{\prime}\rightarrow\pm\infty$.

Let us first consider $x^{\prime}\rightarrow\infty$. According to
Eq.\ (\ref{Green function}) the $x^{\prime}$ dependence of the Green's
function pair $\mathbf{G},\mathbf{\Gamma}$ consist of modes $\mathbf{u}%
_{s}(x^{\prime})^{\ast}$ with $v_{g}^{(s)}<0$ since here $x^{\prime}>x$ for
any finite $x$. On the other hand the field pair $\mathbf{u},\mathbf{T}$ are
made up of the incident wave, and waves scattered from the roughness at finite
$x$, and so consist of modes $\mathbf{u}_{r}(x^{\prime})$ with $v_{g}^{(r)}%
>0$. The integral in Eq.\ (\ref{total field2}) over the section at $x^{\prime
}\rightarrow\infty$ is therefore the sum of terms involving $\int\int\left[
u_{i}^{(s)\ast}T_{ix}^{(r)}-u_{r}^{(r)}T_{ix}^{(s)\ast}\right]  \,dydz$ with
$v_{g}^{(r)}$ and $v_{g}^{(s)}$ of opposite sign. All these terms are zero by
Eq. (\ref{Eq_reciprocity_1}), and so there is no contribution from the section
at $x^{\prime}\rightarrow\infty$.

Similar arguments apply to the section at $x^{\prime}\rightarrow-\infty$. The
Green function is made up of modes with $v_{g}>0$. The scattered component of
the field $\mathbf{u}$ consists of modes with $v_{g}<0$, and there is no
contribution to the integral over the section from these modes. On the other
hand the incident wave $\mathbf{u}^{\mathrm{in}}$ is mode $\mathbf{u}_{m}$
with $v_{g}^{(m)}>0$, and there is the single term with $v_{g}^{(n)}%
=v_{g}^{(m)}$ surviving in the sum over modes in the Green function. Using
Eq.\ (\ref{Eq_reciprocity_2}) the integral just gives $u_{q}^{(m)}(x)$.
Writing $\mathbf{u}=\mathbf{u}^{\mathrm{in}}+\mathbf{u}^{\mathrm{sc}}$ then
leads to Eq.\ (\ref{scatterd-field-w/bd}) in the text.

\section{Energy Flux for Flexural Modes}

\label{App_TPP}The classical thin plate approximation of setting $T_{zi}=0$ is
not sufficient to calculate the energy flux of the flexural modes using the
integral Eq.\ (\ref{Power_integral_ex}). In this appendix we evaluate the
correct expression for the energy flux by two different methods, first using
the extended thin-plate theory of Timoshenko\cite{T} (see also Graff\cite{G}),
and then using a method in terms of the energy of plate deformations\cite{LF7}
that avoids these difficulties.

In the extended thin plate approximation of Timoshenko the $z$-dependence of
the in-plane displacements is still approximated as linear%
\begin{align}
u_{x}(x,y,z)  &  \simeq z\psi_{x}(x,y),\label{Eq_uxuy}\\
u_{y}(x,y,z)  &  \simeq z\psi_{y}(x,y).
\end{align}
However, the $x,y$ dependence is no longer assumed to be given by the gradient
of the mean vertical displacement $\bar{u}_{z}(x,y)$, but by the more general
expression%
\begin{equation}
\mathbf{\psi}=-\mathbf{\nabla}_{\perp}\bar{u}_{z}+\mathbf{\nabla}_{\perp
}S+\mathbf{\nabla}_{\perp}\times(\zeta\hat{z}) \label{Eq_def}%
\end{equation}
introducing the scalar potential $S(x,y)$ and vector potential $\zeta(x,y)$
defining the corrections to the in-plane strain and rotation. Here
$\mathbf{\nabla}_{\perp}=(\partial_{x},\partial_{y})$ is the horizontal
gradient. In addition the vertically averaged stress $T_{zx}$ is taken to be%
\begin{equation}
T_{zx}\simeq\kappa^{2}\frac{E}{2\left(  1+\sigma\right)  }(\partial
_{x}\bar{u}_{z}+\partial_{z}u_{x}) \label{Eq_Tzx_kappa}%
\end{equation}
(with a similar expression for $T_{zy}$ given by replacing the subscript $x$
with $y$ everywhere). Here the ``shear correction factor'' $\kappa$, a number
of order unity, is introduced to take into account deviations of the in-plane
displacements from the assumed linear dependence on $z$\cite{T}. In the usual
thin plate approximation $T_{zi}$ are set to zero and $\psi=-\mathbf{\nabla
}_{\perp}w$, so that $(u_{x},u_{y})=-z\mathbf{\nabla}_{\perp}w$.

With the Timoshenko approximations, the equations of motion for the three
components of displacement are now investigated.

The equations of motion for the horizontal displacement lead to an equation
relating $\mathbf{\psi}$ to $\bar{u}_{z}$\cite{G}%
\begin{equation}
\frac{D}{2}\left\{  (1-\sigma)\nabla^{2}\mathbf{\psi}+(1+\sigma)\mathbf{\nabla
}_{\perp}\mathbf{\nabla}_{\perp}\cdot\mathbf{\psi}\right\}  -\kappa^{2}\mu
d\left(  \mathbf{\psi}+\mathbf{\nabla}_{\perp}\bar{u}_{z}\right)  =0
\label{Psi}%
\end{equation}
(remember $D=Ed^{3}/12(1-\sigma^{2})$, with $E$ Young's modulus,
and $\mu$ the shear modulus). The inertial terms
$\partial_{t}^{2}\mathbf{\psi}$ \ turn out to be negligible in
this equation. Using Eq.\ (\ref{Eq_def}),
Eq.\ (\ref{Psi}) becomes%
\begin{equation}
D\mathbf{\nabla}_{\perp}\nabla_{\perp}^{2}(S-w)-\kappa^{2}\mu d\mathbf{\nabla
}_{\perp}S+\frac{D}{2}(1-\sigma)\mathbf{\nabla}_{\perp}\times(\nabla_{\perp
}^{2}\zeta\hat{z})-\kappa^{2}\mu d\mathbf{\nabla}_{\perp}\times(\zeta\hat
{z})=0. \label{Eq_S_zeta}%
\end{equation}
Taking the vertical curl of Eq.\ (\ref{Eq_S_zeta}) gives%
\begin{equation}
\frac{D}{2}(1-\sigma)\nabla_{\perp}^{2}\Omega-\kappa^{2}\mu d\,\Omega=0
\end{equation}
with $\Omega=\hat{z}\cdot\mathbf{\nabla}_{\perp}\times\mathbf{\psi}%
=-\nabla_{\perp}^{2}\zeta$ the rotation. For a wave disturbance $e^{ikx}$,
this gives an exponential dependence on $y$, $e^{\pm\lambda y}$ with%
\begin{equation}
\lambda^{2}\simeq\frac{2\kappa^{2}\mu d}{D(1-\sigma)}\sim d^{-2}.
\end{equation}
Since $\lambda^{-1}\sim d\ll W$, the rotation will be large only
over a boundary layer region with width of order $d$ near the
edges $y=\pm W/2$,
where the solution takes the form%
\begin{equation}
\Omega(x,y\simeq\pm W/2)\simeq\Omega(\pm W/2)e^{ikx}e^{-\lambda|y\mp W/2|}.
\end{equation}
The vector potential $\zeta$ has a similar solution, so that last two terms in
Eq.\ (\ref{Eq_S_zeta}) cancel. This leaves for the scalar potential $S$%
\begin{equation}
\mathbf{\nabla}_{\perp}\left(  D\nabla_{\perp}^{2}(S-\bar{u}_{z})-\kappa
^{2}\mu dS\right)  =0
\end{equation}
which immediately gives%
\begin{equation}
D\nabla_{\perp}^{2}S-\kappa^{2}\mu dS=D\nabla_{\perp}^{2}\bar{u}_{z}.
\label{Eq_Sw1}%
\end{equation}
(We are only interested in $\mathbf{\nabla}_{\perp}S$ and so do not need to
keep track of the arbitrary gradient-free function that could be added to this equation.)

The equation of motion for the vertical displacement is\cite{G}%
\begin{equation}
\kappa^{2}\mu d\nabla_{\perp}^{2}S=-\rho d\omega^{2}\bar{u}_{z}.
\label{Eq_Sw2}%
\end{equation}
Together Eqs.\ (\ref{Eq_Sw1},\ref{Eq_Sw2}) give%
\begin{equation}
\rho d\omega^{2}(\bar{u}_{z}-\frac{D}{\kappa^{2}\mu d}\nabla_{\perp}%
^{2}\bar{u}_{z})=D\nabla_{\perp}^{4}\bar{u}_{z}.
\end{equation}
This is the usual fourth order wave equation, with a small correction term of
order $(d/W)^{2}$ (the second term in the brackets on the left hand side).
Note that solutions to this equation vary on the long scale of order $k^{-1}$
or $W$, and not the small scale $\lambda^{-1}\sim d$, so that to a good
approximation we have%
\begin{subequations}
\begin{align}
\rho d\omega^{2}\bar{u}_{z}  &  =D\nabla_{\perp}^{4}\bar{u}_{z},\\
S  &  =-\frac{D}{\kappa^{2}\mu d}\nabla_{\perp}^{2}\bar{u}_{z}.
\label{Eq_S_approx}%
\end{align}
The first equation is now the standard fourth order wave equation.
The second equation for $S$ shows it to be small compared with
$\bar{u}_{z}$ by of order $(d/W)^{2}$.

The boundary conditions at the edges are that all stresses are zero, so that
in particular at $y=\pm W/2$%
\end{subequations}
\begin{equation}
\int dzT_{zy}=\kappa^{2}\mu d(\partial_{y}\bar{u}_{z}+\psi_{y})=0.
\end{equation}
Substituting Eq.\ (\ref{Eq_def}) into this gives%
\begin{subequations}
\begin{equation}
\partial_{y}S-\partial_{x}\zeta=0. \label{Eq_bc3}%
\end{equation}
Equation (\ref{Eq_bc3}) together with Eq.\ (\ref{Eq_S_approx}) tells us the
size of the $\zeta$ correction, which at $y=\pm W/2$ takes the value%
\end{subequations}
\begin{equation}
\zeta(x,y=\pm W/2)=-\frac{D}{\kappa^{2}\mu d}\frac{1}{ik}\left.  \left(
\partial_{y}\nabla_{\perp}^{2}\bar{u}_{z}\right)  \right|  _{y=\pm W/2}.
\end{equation}
This expression can be simplified using the boundary condition $T_{yy}=0$ at
$y=\pm W/2$, which from Eq.\ (\ref{Tyy}) and Eqs.\ (\ref{Flex_ux}%
,\ref{Flex_uy}) gives at $y=\pm W/2$%
\begin{equation}
\partial_{y}^{2}\bar{u}_{z}=-\sigma\partial_{x}^{2}\bar{u}_{z}=\sigma
k^{2}\bar{u}_{z},
\end{equation}
so that%
\begin{equation}
\zeta(x,y=\pm W/2)=-\frac{ikD(1-\sigma)}{\kappa^{2}\mu d}\left.  \left(
\partial_{y}\bar{u}_{z}\right)  \right|  _{y=\pm W/2}. \label{Eq_zeta}%
\end{equation}
The potential $\zeta$ is only large in the boundary layers near the edges
where it takes the form%
\begin{equation}
\zeta(x,y\simeq\pm W/2)=-\frac{ikD(1-\sigma)}{\kappa^{2}\mu d}\left.
(\partial_{y}\bar{u}_{z})\right|  _{y=\pm W/2}e^{-\lambda|y\mp W/2|}
\label{zeta-final}%
\end{equation}

Thus finally we have expressions for the horizontal displacement
field, Eqs.\ (\ref{Eq_def}) and (\ref{Eq_uxuy}) together with
Eqs.\ (\ref{Eq_S_approx}) and (\ref{zeta-final}) defining $S$ and
$\zeta$, and Eq.\ (\ref{Eq_Tzx_kappa}). These can be used to
calculate the additional contribution to the energy flux coming
from the $T_{zx}$ term in Eq.\ (\ref{Power_integral_ex}). (The
corrections to $u_{x}$ and $u_{y}$ derived here do not change the
contributions from the first two terms in Eq.\
(\ref{Power_integral_ex}) to the order we require, since these
terms are already third order in the small parameter $d/W$.)

We therefore need to evaluate%
\begin{equation}
\int\int T_{zx}u_{z}^{\ast}dydz\simeq\kappa^{2}\mu d\int dy(\partial
_{x}S\,+\partial_{y}\zeta\,)\bar{u}_{z}^{\ast}. \label{z_flux}%
\end{equation}
Both terms in the integral give contributions at the same order. The first
term, coming from the correction to the in-plane strain
Eq.\ (\ref{Eq_S_approx}), is%
\begin{equation}
\kappa^{2}\mu d\int dy(\partial_{x}S)\,\bar{u}_{z}^{\ast}=-D\int(\partial
_{x}\nabla_{\perp}^{2}\bar{u}_{z})\bar{u}_{z}^{\ast}\,dy.
\end{equation}
The second term in the integrand is only large in the boundary layer region
near the edges and from Eq.\ (\ref{Eq_zeta}) evaluates to the edge
contributions%
\begin{equation}
\kappa^{2}\mu d\int dy\,(\partial_{y}\zeta)\,\bar{u}_{z}^{\ast}=-ikD(1-\sigma
)\left.  [(\partial_{y}\bar{u}_{z})\bar{u}_{z}^{\ast}]\right|  _{y=-W/2}%
^{y=W/2}.
\end{equation}

Combining these expressions for Eq.\ (\ref{z_flux}) with
Eqs.\ (\ref{Poynting-inp}) together with (\ref{Flex_ux}) and (\ref{Flex_uy})
yields the final expression%
\begin{multline}
P\simeq\frac{\omega kD}{2}\operatorname{Re}\left\{  \int dy\left[
2k^{2}\bar{u}_{z}\bar{u}_{z}^{\ast}+\left(  1-\sigma\right)  \left(
\partial_{y}\bar{u}_{z}\right)  \left(  \partial_{y}\bar{u}_{z}\right)
^{\ast}-\left(  1+\sigma\right)  \left(  \partial_{y}^{2}\bar{u}_{z}\right)
\bar{u}_{z}^{\ast}\right]  \right. \label{Timoshenko_power}\\
\left.  +\left[  \left(  1-\sigma\right)  \left(  \partial_{y}\bar{u}_{z}%
\right)  \bar{u}_{z}^{\ast}\right]  _{y=-W/2}^{y=W/2}\right\}  .
\end{multline}
which is identical to Eq.\ (\ref{Poynting-power}).

An alternative approach to calculate the energy flux is to use the expression
for the energy of distortions of the plate evaluated using the lowest order
expressions Eqs.\ (\ref{Txx},\ref{Tyy}) and (\ref{Flex_ux},\ref{Flex_uy}%
)\cite{LF7}%
\begin{equation}
F=\frac{1}{2}D\int\int\left[  \left(  \nabla_{\perp}^{2}\bar{u}_{z}\right)
^{2}+2(1-\sigma)\left\{  \left(  \frac{\partial^{2}\bar{u}_{z}}{\partial
x\partial y}\right)  ^{2}-\frac{\partial^{2}\bar{u}_{z}}{\partial x^{2}%
}\frac{\partial^{2}\bar{u}_{z}}{\partial y^{2}}\right\}  \right]  dxdy.
\end{equation}
It turns out that the higher order corrections discussed above are not needed
in this expression, and so we can derive the energy flux without these
difficulties. The functional derivative of $F$ with respect to $\bar{u}_{z}$
yields the vertical force per unit area in the interior of the plate, which
can be used to derive the fourth order wave equation, as well as expressions
for the energy flux into the plate across the boundaries. The latter
expressions give us the result for the energy flux along the beam%
\begin{equation}
P=\frac{1}{2}\operatorname{Re}\left\{  -i\omega\left[  \int M_{x}\theta
_{x}^{\ast}+V\bar{u}_{z}^{\ast}\,dy+\left.  (F_{c}\bar{u}_{z}^{\ast}\right|
_{y=W/2}\left.  +F_{c}\bar{u}_{z}^{\ast}\right|  _{y=-W/2})\right]  \right\}
\label{flux-CL}%
\end{equation}
where%
\begin{equation}
V=-D\partial_{x}[\partial_{x}^{2}\bar{u}_{z}+\left(  2-\sigma\right)
\partial_{y}^{2}\bar{u}_{z}] \label{Eq_V}%
\end{equation}
is the effective vertical force that couples to the vertical displacement
$\bar{u}_{z}$,%
\begin{equation}
M_{x}=-D\left(  \partial_{x}^{2}\bar{u}_{z}+\sigma\partial_{y}^{2}%
\bar{u}_{z}\right)  \label{Eq_Mx}%
\end{equation}
is the torque that couples to the angular displacement $\theta_{x}%
=\partial\bar{u}_{z}/\partial x$, and%
\begin{equation}
F_{c}(y=\pm W/2)=\pm\left.  2D(1-\sigma)\partial_{xy}^{2}\bar{u}_{z}\right|
_{y=\pm W/2} \label{Eq_Fc}%
\end{equation}
is a vertical force localized at the edges of the plate.

Substituting Eqs.\ (\ref{Eq_V}-\ref{Eq_Fc}) into Eq.\ (\ref{flux-CL}) gives
\begin{multline}
P=\frac{1}{2}\operatorname{Re}\left\{  \frac{\left(  i\omega D\right)  }%
{2}\left[  \int dy\left(  \partial_{x}^{2}\bar{u}_{z}+\sigma\partial_{y}%
^{2}\bar{u}_{z}\right)  \left(  -\partial_{x}\bar{u}_{z}\right)  ^{\ast}+\int
dy[\partial_{x}^{3}\bar{u}_{z}+\left(  2-\sigma\right)  \partial_{x}%
\partial_{y}^{2}\bar{u}_{z}]\bar{u}_{z}^{\ast}\right.  \right.
\label{free-energy-power}\\
\left.  \left.  -\left.  2\left(  1-\sigma\right)  \left(  \partial
_{x}\partial_{y}\bar{u}_{z}\right)  \bar{u}_{z}^{\ast}\right|  _{y=W/2}%
+\left.  2\left(  1-\sigma\right)  \left(  \partial_{x}\partial_{y}%
\bar{u}_{z}\right)  \bar{u}_{z}^{\ast}\right|  _{y=-W/2}\right]  \right\}
\end{multline}
Evaluating $\partial_{x}=ik$, and using integration by parts, we again get
Eq.\ (\ref{Poynting-power}).

\end{document}